\documentclass[a4paper,11pt]{article}
\pdfoutput=1 

\usepackage{jinstpub} 

\title{Measurement of parameters of scintillating bars with wavelength-shifting fibres and silicon photomultiplier 
readout  for the SHiP Muon Detector}

\author[a]{W.~Baldini,}
\author[b]{A.~Blondel,}
\author[c] {A.~Calcaterra,}
\author[d]{R.~Jacobsson,}
\author[e]{A.~Khotjantsev,}
\author[e,f,g] { Yu.~Kudenko,} 
\author[e] {V.~Kurochka,}
\author[c] {G.~Lanfranchi,}
\author[e]{A.~Mefodiev,}
\author[e]{O.~Mineev,}
\author[h]{A.~Montanari,}
\author[b]{E.~Noah~Messomo,}
\author[c] {A.~Saputi,}
\author[h] {N.~Tosi}

\affiliation[a]{INFN - Sezione di Ferrara, via Saragat 1, 44122 Ferrara, Italy}
\affiliation[b]{DPNC, Section de Physique, Universit$\acute{e}$ de Gen$\grave{e}$ve, Geneva, Switzerland}
\affiliation[c]{INFN - Laboratori Nazionali di Frascati, via E. Fermi 40, 00044 Frascati (Rome), Italy}
\affiliation[d] {European Organization for Nuclear Research (CERN), Geneva, Switzerland}

\affiliation[e]{Institute for Nuclear Research of the Russian Academy of Science,  \\
 pr. 60-letiya Oktyabrya 7a, Moscow, Russia 117312}
\affiliation[f]{Moscow Institute of Physics and Technology,  Institutskiy per. 9, Dolgoprudny, Moscow region, Russia, 141701}
\affiliation[g]{National Research Nuclear University MEPhI,  Kashirskoe sh. 31, Moscow,
Russia, 115409}
\affiliation[h]{INFN - Sezione di Bologna,Viale Berti Pichat, 6/2, 40127 Bologna, Italy}

\emailAdd{Gaia.Lanfranchi@lnf.infn.it}

\abstract{The light yield and the time resolution of different types of 3 m long  scintillating bars  instrumented with wavelength 
shifting fibres and read out by different models of silicon photomultipliers have been measured at a test 
beam at the T9 area at the CERN Proton Synchrotron.  
The results obtained with different configurations are presented.  
A time resolution better than 800~ps, constant along the bar length within 20\%,
and a light yield of $\sim 140 \;(70) $ photoelectrons are obtained for bars 3 m long, $\sim$4.5 (5) cm wide and 2 (0.7) cm thick.
These results  nicely match the requirements for the Muon Detector of the SHiP experiment.}


\arxivnumber{1612.01125} 

\begin{document}
\maketitle
\flushbottom

\section{Introduction}
\label{sec:intro}
Extruded plastic scintillator bars with wavelength shifting (WLS) fibres and Silicon Photomultiplier (SiPM) readout
are considered an established technology for massive tracking calorimeters in long-baseline neutrino oscillation experiments. 
The MINOS experiment~\cite{minos} employes extruded
bars of $(1\times 4.1 \times 800)$ cm$^3$ size with 9 m long WLS fibres. A fine-grained detector in the Miner$\nu$a
experiment~\cite{Minerva} is made of triangular-shaped 3.5 m long strips and WLS fibres of 1.2 mm
diameter. Other experiments that use the same technology are Belle II~\cite{Belle2} and T2K~\cite{T2K}.
This technology has been considered also as a viable option for the muon system of the SHiP experiment~\cite{ship} 
proposed at the CERN SPS. 
The SHiP muon detector  comprises four muon stations interleaved by iron filters, each station with a 
transverse dimension of $(6 \times 12)$ m$^2$ for  a total active area of 288 m$^2$. Each station has to provide both spatial
and time information. The $x,y$ coordinate will be obtained by the crossing of horizontal and vertical bars 3 m long, with a granularity
to be defined between 5 and 10 cm. The time information is provided by the average of the times measured at both ends of the bars.
A time resolution better than $1$ ns per station is required.
This paper shows the results  obtained on different types of extruded scintillating bars instrumented with 
different types of WLS fibres and SiPMs 
measured at a test beam held at the T9 area of the CERN Proton Synchrotron (PS) in the period October 14-28, 2015.

\section{The prototypes}
\label{sec:prototypes}
Given the relatively large area, a good choice for the SHiP Muon system is the rather inexpensive scintillators
produced at the FNAL-NICADD facility~\cite{fnal1, fnal2} which are fabricated by co-extrusion with a thin
layer of $TiO_2$ around the active core. Another possibility is the polystyrene scintillator bars
extruded at UNIPLAST plant (Vladimir, Russia)~\cite{uniplast}. 

Since the attenuation length of the plastic scintillator is rather short, 
the light produced by the particle interaction has to be collected, re-emitted, and transported to the photodetectors
efficiently by WLS fibres. These fibres need to have a good light yield to ensure a high detection
efficiency for fibre lengths of $\sim$ 3 m. Possible choices for WLS fibres are those produced by
Saint-Gobain~\cite{saint-gobain} and from Kuraray~\cite{kuraray} factories. Both companies
produce multiclad fibres with long attenuation length ($\sim$4 m) and good trapping efficiency
($\sim$ 5\%). The fibres from Kuraray have a higher light yield 
while Saint-Gobain fibres have a faster response ($\sim$2.7 ns versus $\sim$10 ns of the Kuraray), which ensures
a better time resolution for the same light yield.

Scintillating bars from NICADD and UNIPLAST companies of different lengths, widths and thicknesses
were instrumented with different types and numbers of WLS fibres from Kuraray and 
Saint-Gobain manufacturers and read out   by different types of SiPMs from Hamamatsu and AdvanSiD (FBK) companies.


Table~\ref{tab:prototypes1} and Table~\ref{tab:prototypes2} show the main parameters of scintillating bars from NICADD and UNIPLAST 
manufacturers, respectively.

\begin{table}[htbp]
\centering
\caption{\label{tab:prototypes1} Prototypes of extruded scintillator bars from NICADD manufacturer. All the bars were instrumented
with fibres Kuraray WLS Y11(200) S-type except the S2 bar that has been instrumented with fibres from the Saint Gobain company (BCF92). 
The fibres in the L1, L2 and L4 bars were read out at both ends. The fibres in the S1, S2, S5 and S8 bars were read out only at one end.
The main parameters of the photosensors are shown in Table~\ref{tab:sipm}.}
\smallskip
\begin{tabular}{|l|c|c|c|c|}
\hline
              &  Bar dimensions                                      & number of fibres/bar & fibre diameter & SiPM model \\ 
                                  & (h $\times$ w $\times$ l) mm$^3$        &    &  [mm] &  (AdvanSiD company)  \\ \hline
L1          &  ($10\times 45 \times 3000$) mm$^3$ &   1 fibre in 1 groove      &  2            & ASD-NUV3S-P \\
L2          &  ($20\times 40 \times 3000$) mm$^3$ &  1 fibre in 1 groove      &  2              & ASD-NUV3S-P \\
L4          &  ($20\times 40 \times 3000$) mm$^3$ &   1 fibre in 1 groove      &  1.2           & ASD-NUV1S-P  \\ \hline
S1          &   ($10\times 45 \times 250$) mm$^3$  &   2 fibres in 1 groove   & 1.2              & ASD-NUV3S-P \\
S2          &   ($10\times 45 \times 250$) mm$^3$  &  2 fibres in 1 groove   & 1.2               & ASD-NUV3S-P \\
S5          &   ($20\times 40 \times 250$) mm$^3$  &  2 fibres in 1 groove   & 1.2              & ASD-NUV3S-P \\
S8          &   ($20\times 40 \times 250$) mm$^3$  &  1 fibre in 1 hole        &  2                   & ASD-NUV3S-P \\ 
\hline
\end{tabular}
\end{table}

\begin{table}[htbp]
\centering
\caption{\label{tab:prototypes2} Prototypes of extruded scintillating bars from UNIPLAST manufacturer. All the bars were instrumented
with fibres Kuraray WLS Y11(200) S-type. In the bars U1, U2 and U3 the fibres were read out at both ends, 
in the U4 bar the two fibres were read out just at one end, opposite with respect to each other. 
The main parameters of the photosensors are shown in Table~\ref{tab:sipm}.}
\smallskip
\begin{tabular}{|l|c|c|c|c|}
\hline
              &  Bar dimensions                                      & number of fibres/bar & fibre diameter & SiPM model \\ 
                                  & (h $\times$ w $\times$ l) mm$^3$        &    & [mm] &   (Hamamatsu company) \\ \hline

U1         &    ($ 7 \times 30 \times 3000$) mm$^3$ & 1 fibre in 1 groove         & 1        & MPPC S13081-050CS \\
U2         &    ($ 7 \times 50 \times 3000$) mm$^3$ & 1 fibre in 1 groove         & 1        & MPPC S13081-050CS \\
U3         &    ($ 7 \times 100 \times 3000$) mm$^3$ & 2 fibres in 2 grooves    & 1        & MPPC S13081-050CS \\
U4         &    ($ 7 \times 100 \times 3000$) mm$^3$ & 2 fibres in 2 grooves    & 1        & MPPC S13081-050CS \\
\hline
\end{tabular}
\end{table}

The L1, L2 and L4 prototypes from NICADD company are 3 m long, of different widths and thicknesses 
as shown in Table~\ref{tab:prototypes1}.
These bars were machined with a single straight groove, on the top face,  to host
1.2 mm or 2 mm diameter Kuraray Y11 (S300) fibres.
The S1,S2,S5 and S8 are 25 cm long bars, of different widths and thicknesses, and were used to test different configurations 
with two fibres hosted in the same groove or one fibre hosted in a hole machined at the center of the bar.
These bars were read out only at one end. The fibres were all fixed with BC-600 optical cement from Saint-Gobain company.
An adhesive aluminum tape has been applied on top of the grooves to reflect the light emerging from the groove. 
The photosensors used for NICADD bars are ASD-NUV3S-P or ASD-NUV1S-P from the Advansid company~\cite{advansid} whose main parameters are listed in 
Table~\ref{tab:sipm}.

\begin{figure}[htb]
\begin{center}
\includegraphics[width=13cm,angle=0]{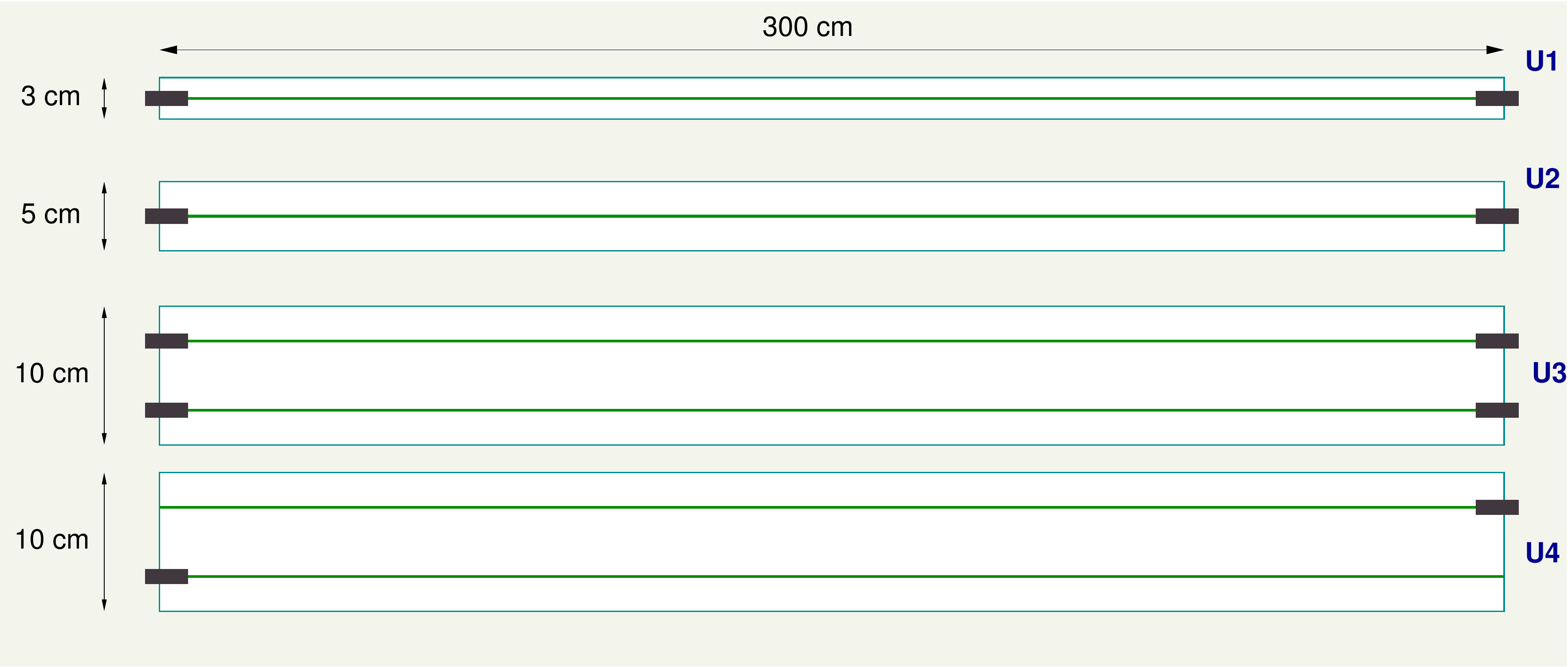}
\caption{\label{fig:bars}  Four types of bars  manufactured at UNIPLAST factory.}
\end{center}
\end{figure}

The 3 m long U1, U2, U3 and U4 prototypes of 0.7 cm thickness
were extruded at the UNIPLAST Factory (Vladimir, Russia) and then   cut to
the  3, 5 and 10 cm wide bars.
The scintillator composition is a polystyrene doped with 1.5\% of paraterphenyl (PTP) 
and 0.01\% of POPOP. The bars were covered by a chemical reflector by etching the scintillator surface  in a chemical 
agent that results in the formation of a white micropore deposit over a polystyrene~\cite{Kudenko:2001qj}.  
The chemical coating is an excellent reflector, besides it dissolves  rough surface acquired during the cutting process.
A 2~mm deep and 1.1~mm wide groove has been machined along a bar central line in 3 and 5 cm 
wide bars to accomodate a WLS  fibre.  The 10 cm wide bars have two grooves running 5 cm apart.
The fibers of all prototypes are read out at both ends except  for the U4 bar where each fiber is read out only at one end.
The four prototypes of the UNIPLAST bars are sketched in Figure~\ref{fig:bars}.

The fibres of one of the 10 cm wide bars  (U4) have been read out only at one end in such a way that the bar is viewed from both ends 
only by two photosensors, one per fibre.  The fibres used for UNIPLAST bars are  Kuraray WLS  Y11 multi-clad fibres of 1~mm diameter.  
The glue used to couple the fibres with a scintillator is optical cement EJ500 from Eljen Technology~\cite{cemento}. 
The same glue has been used to embed optical connectors into the groove.
The plastic optical connector  consists of two parts: a ferrule part glued into the scintillator  and a 
container to hold a Hamamatsu MPPC SiPM.  Both parts are latched by a simple snap-like mechanism. 
A foam spring inside the container provides reliable optical contact between the photosensor and the fibre end.

\section{The photosensors}
\label{sec:sipm}
Bars from NICADD and UNIPLAST manufacturers are read out by photosensors from Advansid and Hamamatsu companies, respectively.
The main parameters of the photosensors are shown in Table~\ref{tab:sipm}.
UNIPLAST bars were instrumented with low crosstalk  Hamamatsu MPPC S13081-050CS~\cite{mppc} 
with sensitive area size of (1.3$\times$1.3)~mm$^2$.  NICADD bars were instrumented with ASD-NUV3S-P~\cite{asd3} and ASD-NUV1S-P~\cite{asd1}
with squared  area of dimensions (3$\times$3)~mm$^2$ and circular area of 1.2 mm diameter, respectively.

%
 %
%
%

\begin{table}[htbp]
\caption{Parameters of SiPMs from different manufacturers.} 
\label{tab:sipm}  
\vspace{.1cm}
\begin{center}
\begin{small}
\begin{tabular}{ccc}
\hline
 Parameter        & Hamamatsu        & AdvanSiD  \\
                        & MPPC                   & ASD \\
                        &S13081-050CS     &  -NUV3S-P   \\
\hline
Pixel size, $\mu$m                & 50                  &  40 \\
Number of pixels                   & 667        &     5520 \\ 
Sensitive area, mm$^2$       &  $1.3 \times 1.3$ &  $3.0 \times 3.0$ \\ 
Gain                                     & 1.5$\times 10^6$  & $2.6 \times 10^6$ (at +4 V over-voltage) \\
Dark rate, kHz/mm$^2$       & $\sim 90 $     &    $< 100$  \\ 
(at T=300 K)                        &                      &  \\
Crosstalk, \%                        & $\sim 1$     & $\sim 25$ (at +4 V over-voltage) \\ 
 PDE                                    & $ \sim 33 \%$ at 520 nm    &   25\% at 520 nm \\ 

Voltage bias, V                     & 70 V at T =  23 $^{\circ}C$  & 30 V at T =  23 $^{\circ}C$   \\ 
\hline
\end{tabular}
\end{small}  
\end{center}
\end{table}

\section{The beam line}
\label{sec:beamline}
The T9 beam at the CERN PS is a secondary beam line produced from a 24 GeV/c primary proton beam
slowly extracted from the PS. The line transports either positive or negative particles 
in the momentum range between 0.5 and 10 GeV/c and with a momentum
resolution of $\sim 0.5\%$.
The beam is a mixed particle beam. Depending on the beam momentum and charge chosen
there are pions, (anti)-protons, $e^+$ or $e^-$ and, at the percent level, also kaons and muons.
For the negatively charged beam, the fraction of electrons can be as high as 80\% for $p=0.5$~GeV/c but
drops to ~5\% at 5 GeV/c, for the ``electron-enriched'' target and to few per mille when the ``hadron-enriched'' target is used.
The maximum particle rate per burst of $10^6$ is achieved for a $p=10$~GeV/c positive beam. 
For negative beams the rates are typically 2-3 times lower and drop significantly at lower
energy. The beam is delivered uniformly over a burst of 0.4 seconds. Depending on scheduling
such a burst is provided typically once or twice every $\sim 15$ seconds.

A negative charged beam with momentum of 10 GeV/c produced with a ``hadron-enriched'' target 
has been used for the measurements discussed in this paper. This choice allows us to have a beam dominated by minimum ionizing particles (mip) and to
minimize the fraction of electrons showering in the material of the  experimental setup. A trigger rate of $\cal{O}$(100 Hz) has been obtained by closing the 
beam collimators, in order to maximize the fraction of single-hit events.

\section{The experimental setup}
\label{sec:setup} 

Bars from NICADD and Vladimir companies were tested simultaneously using  
a common trigger made of two scintillators of $(13 \times 5 \times 1)$~cm$^3$ dimensions
read out by photomultipliers, put in cross one in front and the other behind the bars, and selecting an active 
rectangular area of $(1 \times 5)$~cm$^2$. 
Bars from different companies were read out by independent,  similar setups as described below.
In both setups the signals were read out by digitizers which  record the full
signal waveforms.  This information will be also used for design of the  front-end electronics (FEE) for the muon system of the SHiP experiment.

The experimental setup for NICADD bars is described in the following paragraph.
The coincidence of the two scintillators has been used to start the readout of a buffer of a 10-bit, 8 (4) 
channels, 1 (2) GS/s VME digitizer CAEN V1751.
One of the two trigger scintillator signals has been sent to the digitizer  for time reference.
Signals from the SiPMs were sent via 4 m long RG-174 cables to a 8-channels, 350~MHz bandwidth, 20~db gain, custom 
preamplifier board  based on AD8000 current feedback operational amplifier and then to the digitizer. 
A VME interface has been used to send the data to a PC in the control room via a 30 m long optical fibre.

The signal charge has been measured by integrating the signal waveform within 350~ns window. 
In order to express the light yield in number of photoelectrons (p.e.),
the integrated charge spectra corresponding to dark noise events were registered on a scope and
used to extract the calibration constants for SiPMs. In fact, thanks to the high level of crosstalk ($\sim 25\%$)  of the SiPM from the AdvanSiD company, 
up to  three single photoelectron peaks were clearly visible even in dark noise events and used to evaluate
the charge corresponding to one photoelectron. 
A global uncertainty of 3\% 
has been associated to the calibration constants, taking into account the fitting 
procedure and the effect for temperature fluctuations (roughly 5 $^{\circ}$C day/night).
The optical crosstalk in the SiPMs has been statistically subtracted from the measured light yield.



\begin{figure}[htb]
\begin{center}
\includegraphics[width=0.5\textwidth]{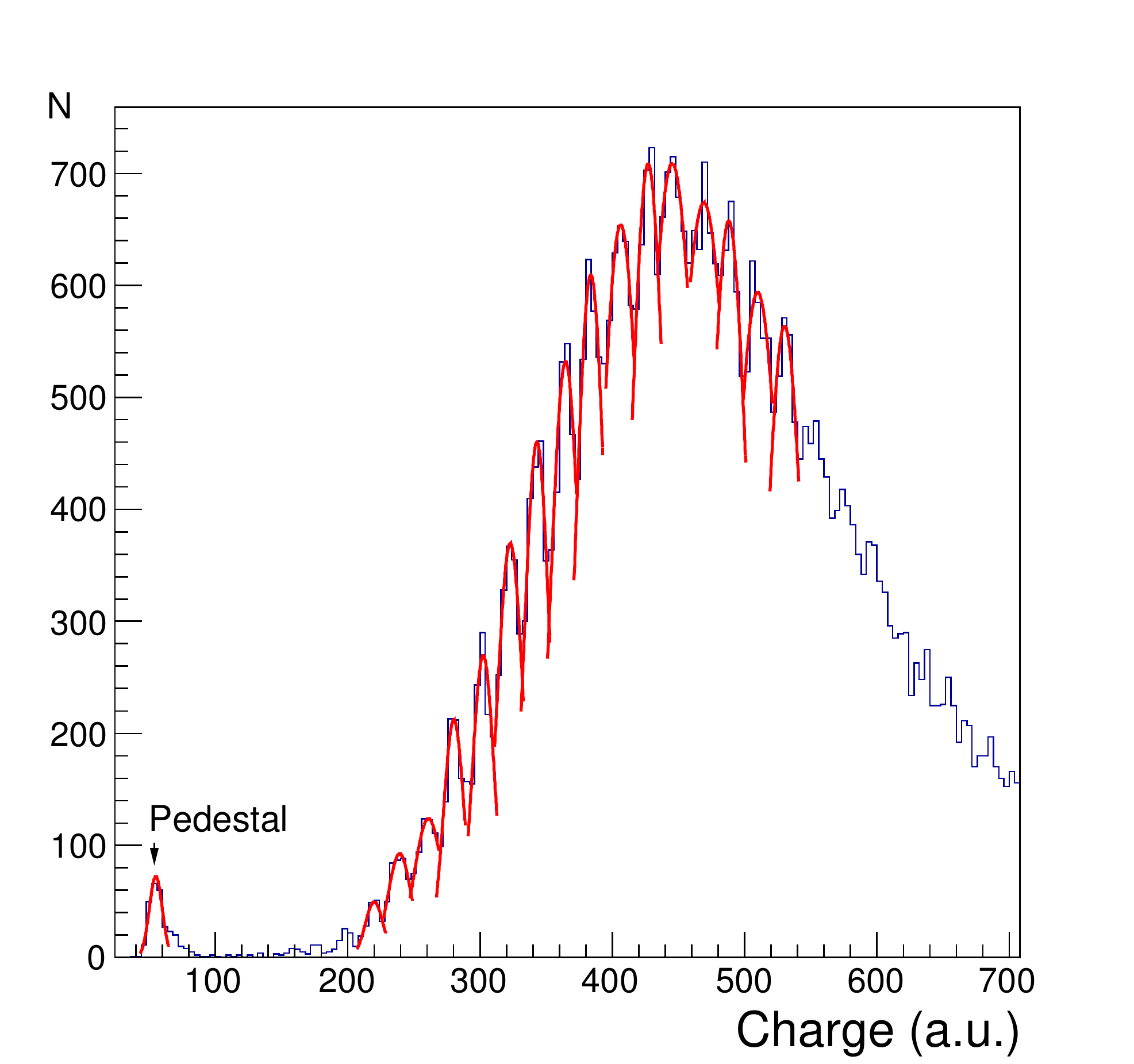}
\caption{Example of MPPC spectrum used for the calibration of the light yield.\label{fig:calib_spectrum}}
\end{center}
\end{figure}

\vskip 2mm
For UNIPLAST bars, signals from the MPPCs were sent through a 2.5~m long 
twisted pair cable to a multi-channel custom made preamplifier with differential inputs. 
The differential inputs suppressed the electronic pickup noise in the experimental hall, nevertheless 
an additional screening  has been required to obtain good separation between single photoelectron (p.e.) 
peaks in the MPPC charge spectra. After shielding  the twisted pair wires with Al-foil connected to ground, 
up to 20 p.e. peaks were visible in the charge spectrum. An example of the spectrum used for the calibration of the light yield
is shown in Figure~\ref{fig:calib_spectrum}.

Then the signals were digitized  by a  12-bit 5-GS/s switched capacitor desktop waveform digitizer CAEN DT5742.  
Six readout channels were operated simultaneously.
The signal charge has been calculated by integrating the signal waveform within 200~ns window. 
The pulse rise time has been analyzed to obtain the timing parameters. 
The calibration of the MPPCs  to express the  light yield in number of p.e. has been done in the 
position where the beam hits a bar  at the far end from the considered MPPC.  The light output in this configuration was around 
20 p.e. so 16 single p.e. peaks were averaged to obtain the calibration coefficients.
The calibration coefficients were measured once. No corrections for temperature fluctuations 
(within roughly 5$^{\circ}$~C day/night) were made. This  contributes to the systematic uncertainty
of the light yield measured from data collected over a few days. The specified value of optical crosstalk in the Hamamatsu 
MPPCs is about 1\% and this factor has been neglected in the light yield determination.

\section{Results}
\label{sec:results}

\subsection{Light yield and attenuation length}
\label{ssec:light-yield}
\begin{figure}[htbp]
\centering 
\includegraphics[width=.7\textwidth]{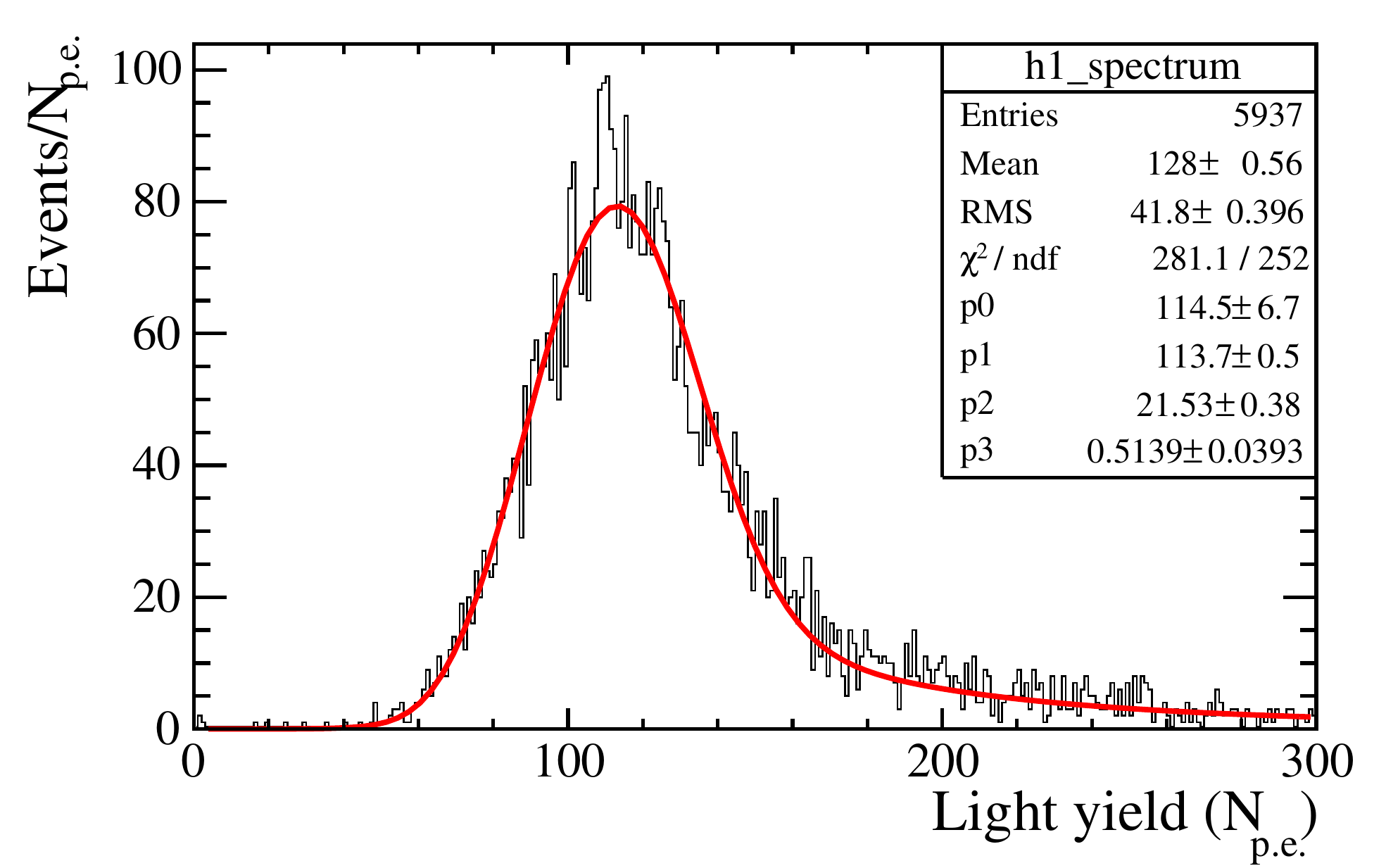}
\qquad
\caption{\label{fig:spectrum} Example of distribution of  the sum of the light yield collected at both ends 
of a 2 cm thick NICADD bar.}
\end{figure}

The light yield for NICADD bars has been obtained by measuring the light yields at both ends
of 3 m long bars and at one end for 25 cm long bars.
An example of the light yield distribution is shown in Figure~\ref{fig:spectrum}. 
The spectrum is fitted with a Gaussian plus a Landau function
with common mean and sigma values. The mean value and its uncertainty obtained from the fit 
are used to determine the light yield and its uncertainty for a given beam position.

\begin{figure}[htbp]
\centering 
\includegraphics[width=.6\textwidth]{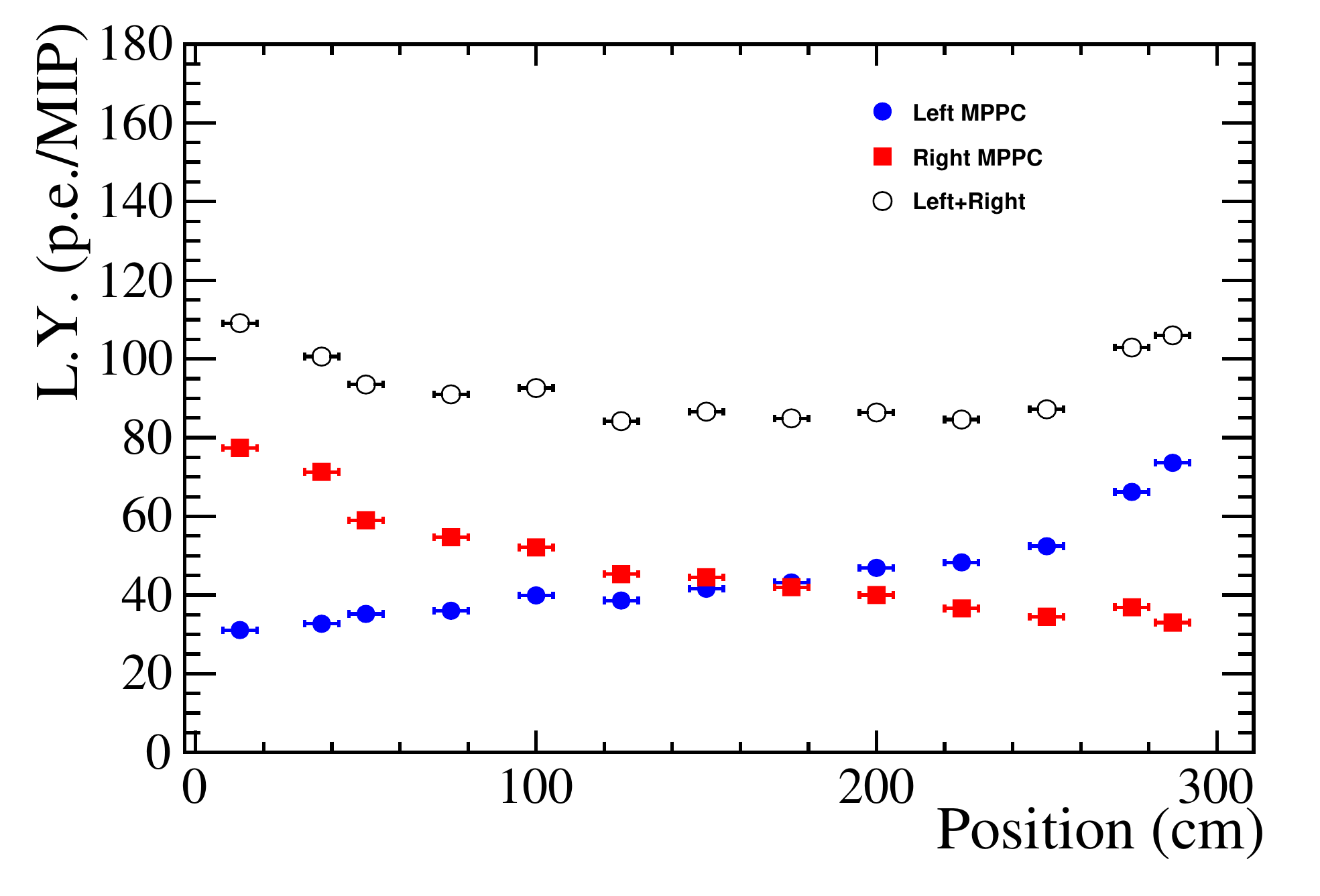}
\qquad
\caption{\label{fig:l1} L1 bar: light yield measured at each end (red squares and blue solid circles) and the sum of the light yields at both ends 
(black open circles)  as a function of the incident beam position along the bar. }
\end{figure}

\begin{figure}[htbp]
\centering 
\includegraphics[width=.6\textwidth]{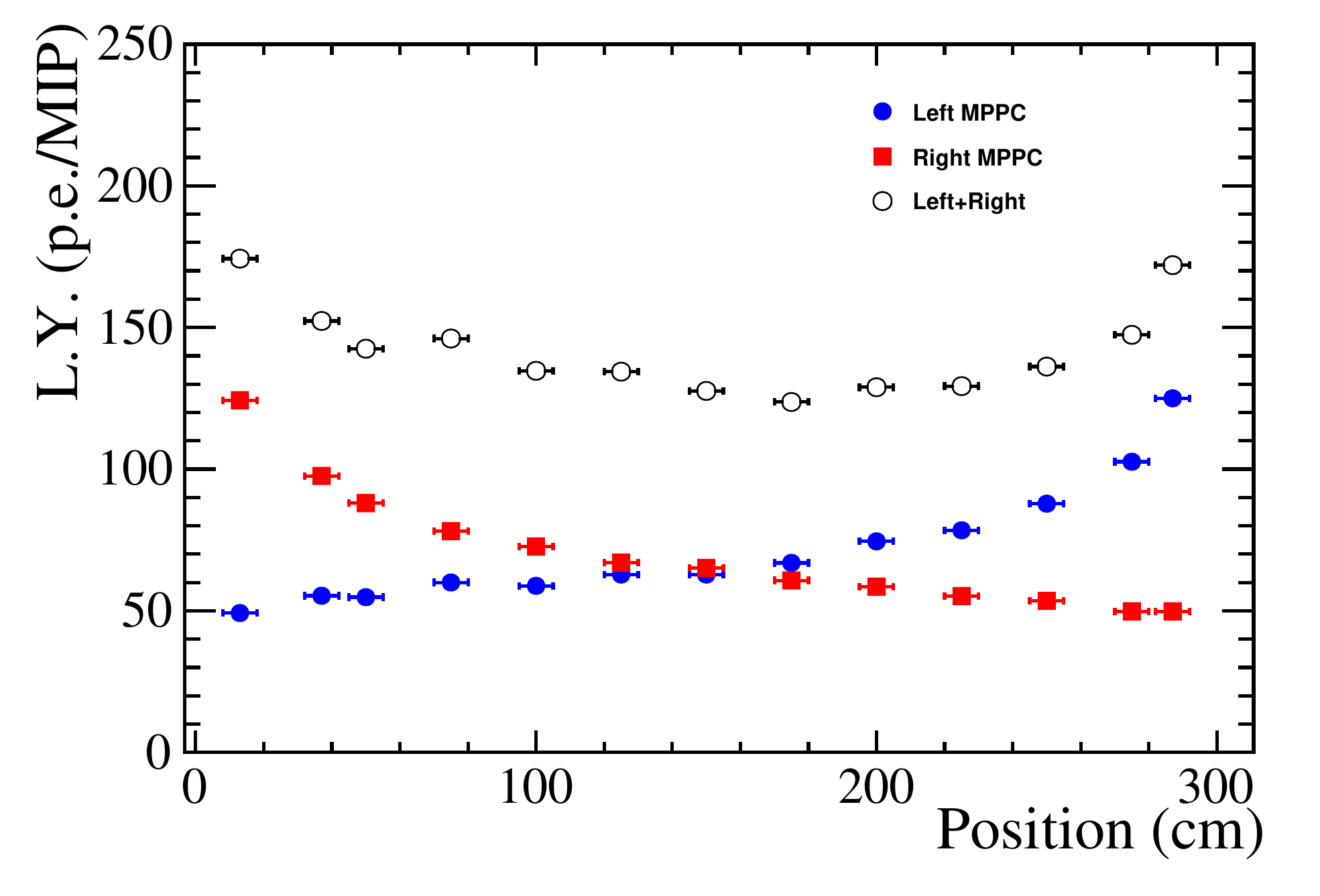}
\qquad
\caption{\label{fig:l2} L2 bar: light yield measured at each end (red squares and blue solid circles) and the sum of the light yields measured at both ends 
(black open circles)  as a function of the incident beam position along the bar.}
\end{figure}

\begin{figure}[htbp]
\centering 
\includegraphics[width=.6\textwidth]{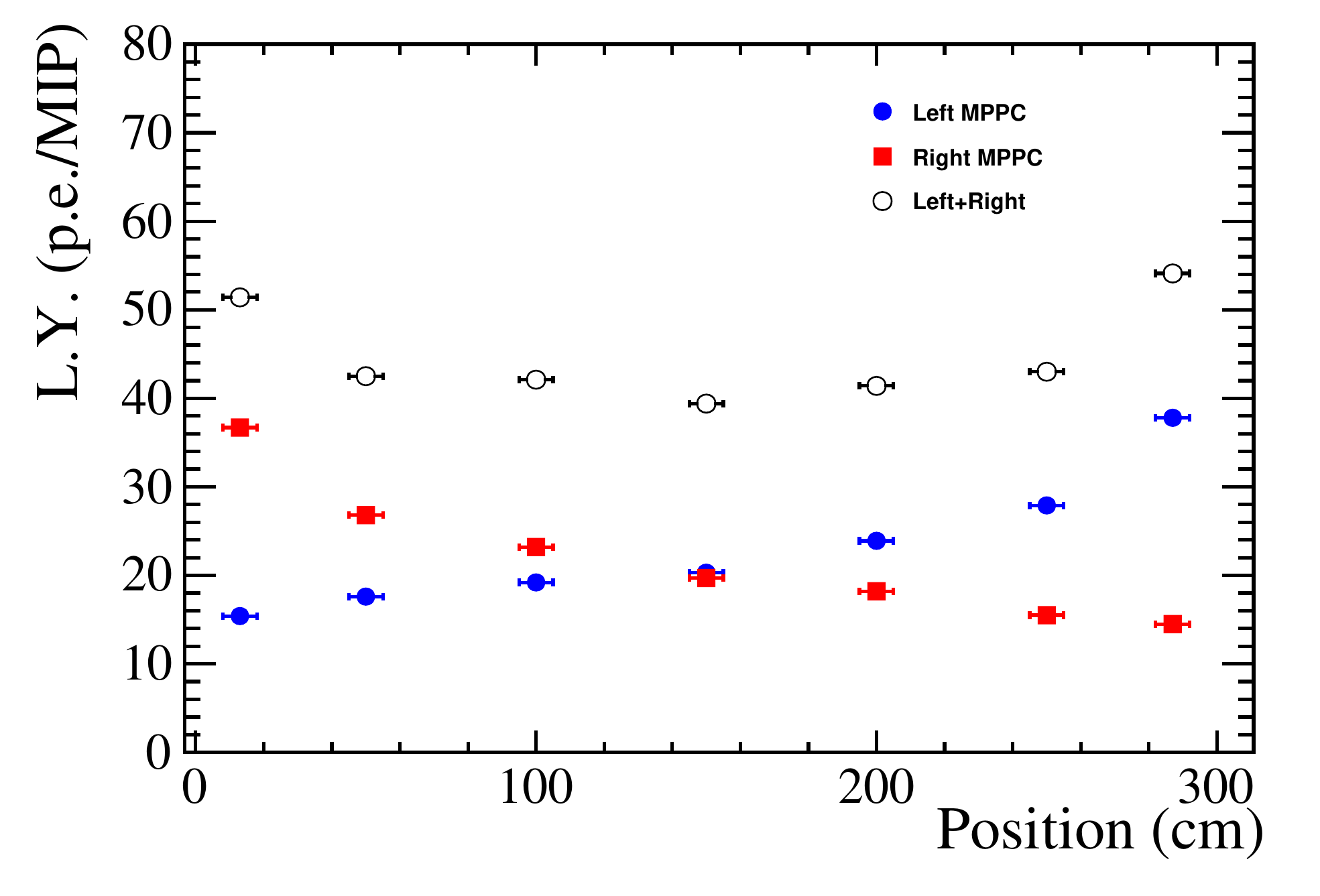}
\qquad
\caption{\label{fig:l4} L4 bar: light yield measured at each end (red squares and blue solid circles) and the sum of the light yields measured at both ends 
(black open circles)  as a function of the beam position along the bar.}
\end{figure}

For long bars, the attenuation of the light during the propagation along the fibre has been determined by measuring the light yield
as a function of the distance of the beam from each photosensor. 
To perform this measurement the bar has been moved with respect to the trigger position by 25 cm steps. 
The results are shown in Figures~\ref{fig:l1},~\ref{fig:l2} and \ref{fig:l4} 
for bars L1, L2, and L4, respectively. 
The attenuation behaviour of the Y11 fibre shows two components, an initial strong attenuation over a distance of about $\sim$25 cm, 
probably dominated by the absorption in the fibre cladding, followed by a much longer attenuation length ($\lambda \sim 4.5-5$ m). 
This is consistent with previously published data~\cite{kudenko}.

\begin{figure}[htbp]
\centering 
\includegraphics[width=.6\textwidth]{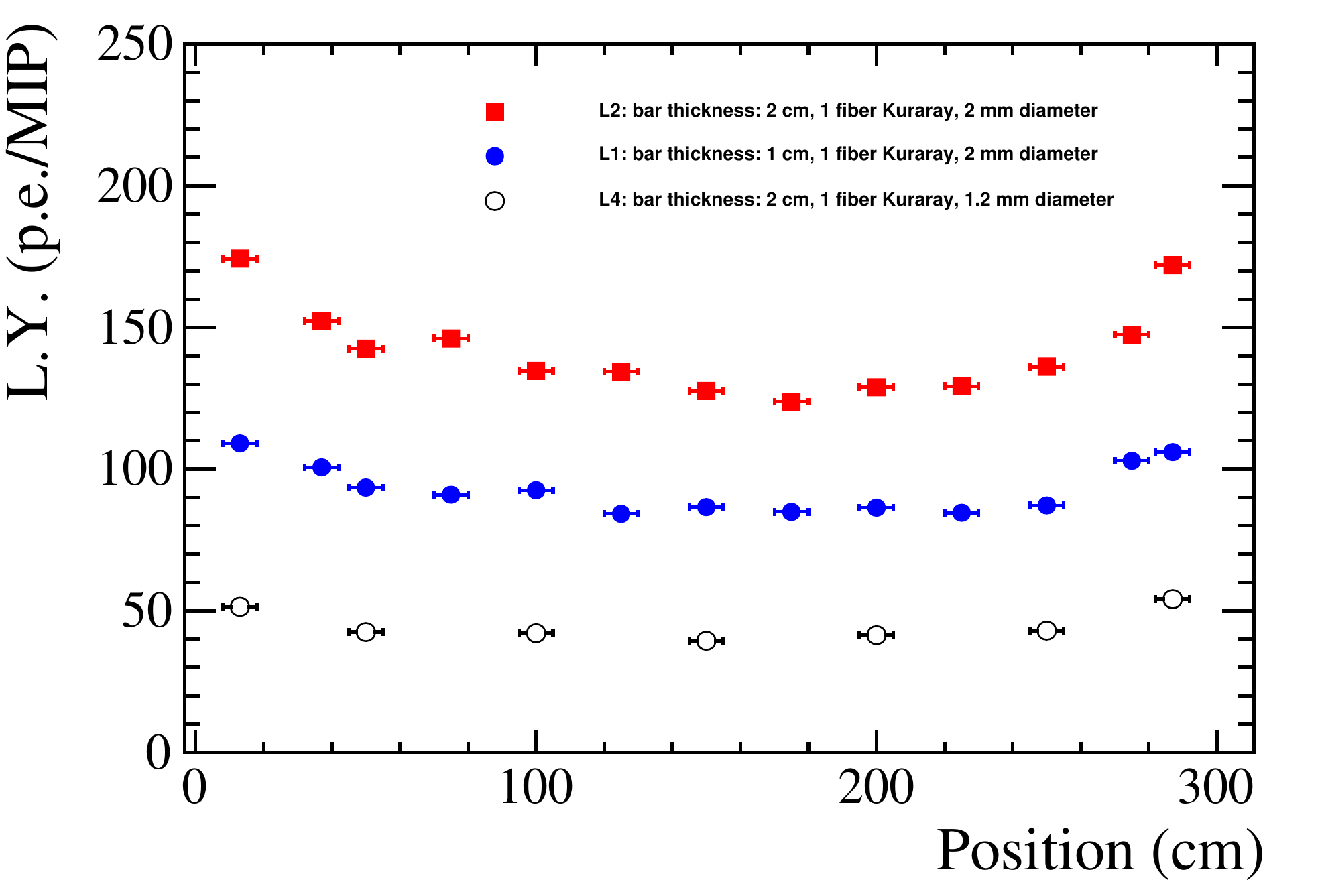}
\qquad
\caption{\label{fig:l124} Sum of the light yield measured at each as a function of the beam position along the bar for L2, L1, L4 bar.}
\end{figure}

The total light yield  measured at both ends is constant within 20\%  along the bar.
Figure~\ref{fig:l124} shows the total light yield  for the  three long bars from the NICADD company.

Table~\ref{tab:npe} shows the light yield measured at one end of the short bars S1,S2,S5 and S8 (as defined in Table~\ref{tab:prototypes1}) when the beam 
impinges at $\sim$ 13 cm far from  the SiPM. As comparison, the light yield measured at one end for long bars is also shown for the same beam position.
The highest light yield is measured for the S5 bar, which is 2 cm thick with two Kuraray fibres,  1.2 mm diameter each, embedded in the same groove.
The comparison of the results obtained with S1 and S2 bars shows that the Bicron fibres emit about half of the light produced by Kuraray fibres with the same
diameter. Another interesting result is that a fibre glued in a groove on the top face of the scintillator bar (L2) produces the same amount of light
of the same fibre glued in a hole in the middle of the bar (S8). In general, the light collected and re-emitted by a fibre is proportional to the transverse area
of the fibre itself (we measured about a factor three more light produced by a  fibre of 2 mm diameter (L2) with respect to a fibre of 
1.2 mm diameter (L4)) while the light yield obtained by doubling the thickness of the scintillator is only 30\% more (L2 versus L1).

\begin{table}[htbp]
\centering
\caption{\label{tab:npe} Light yield measured at one end of the short bars S1, S2, S5 and S8 defined in Table~\ref{tab:prototypes1} when the beam 
impinges at $\sim$ 13 cm far from the SiPM. For comparison of different prototypes, 
the light yield measured at one end for long bars is also shown for the same beam position. The photosensors used for S1, S2, S5, S8, L1, L2, L4 
are different to those used for U1,U2,U3,U4, as shown in Table~\ref{tab:sipm}.
The higher light yield is measured for the S5 bar. The uncertainty is dominated by the systematic one.}
\smallskip
\begin{tabular}{|l|c|}
\hline
             &  light yield [p.e./MIP]   \\ \hline
S1         & $78.0 \pm 2.3$ \\
S2         & $41.0 \pm 1.2$ \\
S5         & $133.0 \pm 4.0$ \\
S8         & $105.9 \pm 3.2$  \\ \hline
L1         & $77.4 \pm 2.3$    \\ 
L2         &  $114.3 \pm 3.4$ \\
L4         &  $36.7 \pm 1.1$  \\ \hline
U1         &  $50.7 \pm 1.2$ \\
U2         &  $45.3 \pm 1.2$ \\
U3         &   $ 22.3 \pm 0.6$ \\
U4         &   $ 19.8 \pm 0.6$ \\ \hline
\hline
\end{tabular}
\end{table}

\begin{figure}[htb]
\begin{center}
\includegraphics[width=15cm,angle=0]{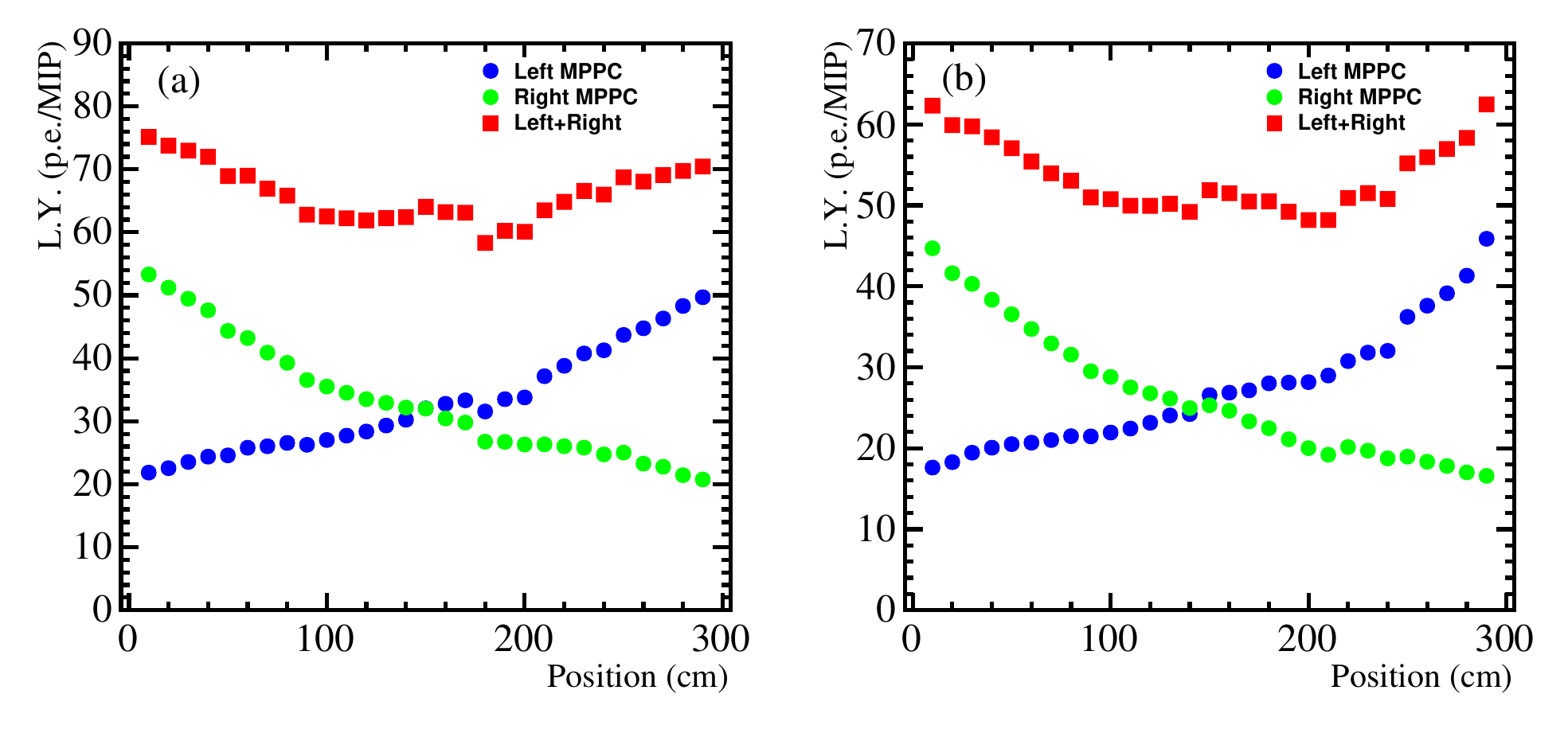}
\caption{\label{fig:ly35} Light yield scan for 3 cm bars U1 (a) and 5 cm bars U2 (b).}
\end{center}
\end{figure}

\vskip 2mm
The light yield for UNIPLAST bars has been measured by using three samples of 3 cm wide bars (U1), three samples of 5 cm wide bars (U2) 
and single prototypes  of 10 cm wide bars (U3 and U4). Figure~\ref{fig:ly35}  shows the average result for 
all three tested bars of the same width. 
The position of the bar with respect to the beam has been changed in steps 10 cm each, collecting 29 points altogether.  

Scan results are shown in Figure~\ref{fig:ly35} for 3 cm (U1) and 5 cm (U2) wide bars. The total light yield from 
both ends has been measured to be about 60 and 50 p.e. per minimum ionizing particle (MIP) for U1 and U2 respectively,
when the beam impinges at the center of the bars. The light yield is higher by 20\% when the beam impinges near the ends.

\begin{figure}[htb]
\begin{center}
\includegraphics[width=15cm,angle=0]{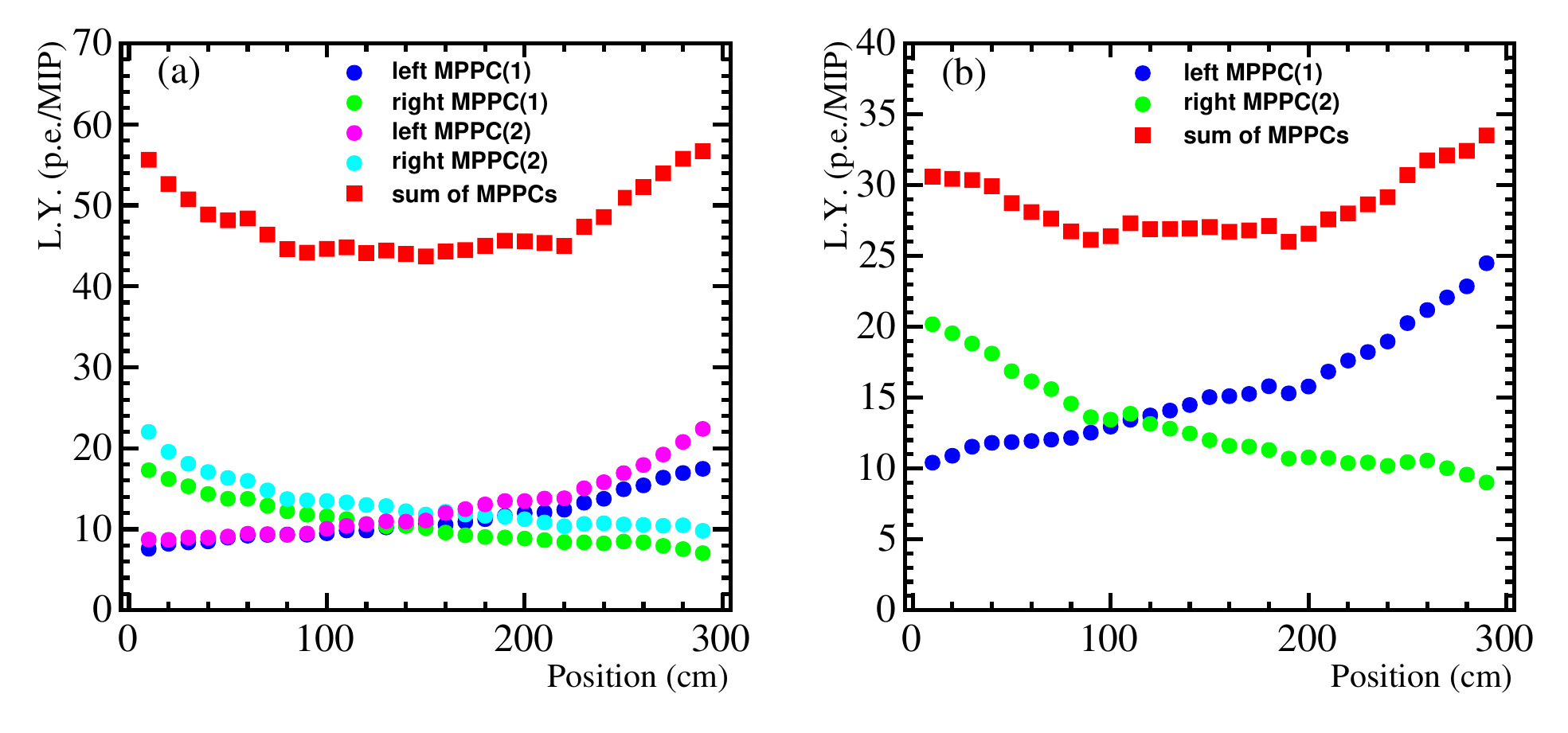}
\caption{\label{fig:ly10} Light yield scan for 10 cm bars U3 (a) and  U4 (b).}
\end{center}
\end{figure}

Figure~\ref{fig:ly10} shows the scan results for 10 cm wide bars, U3 type (left) and U4 (right), respectively.
For U3 bars, the total light yield from all 4 MPPCs
 is about 45 p.e./MIP when the beam impinges at the center of the bars.  Plot on the right shows  the light yield for the bar U4, 
where two WLS fibres are read out by two MPPCs, one per fibre at opposite bar ends. 
This configuration gives the total light output of about 27 p.e./MIP in the center.

The scan has been also performed in the transverse dimension of the bar, to investigate how the light yield depends in the distance between the hit point and the WLS fibre.
This scan was done with the beam impinging in the middle of the bar length.
Bars 3, 5 and 10 cm wide were tested, with a single fibre read out at both ends.
Only events selected by a  ($3\times 3$)~mm$^2$ area small plastic counter read out by a MPPC 
in the same way as the bars under test were considered.  

\begin{figure}[htb]
\begin{center}
\includegraphics[width=15cm,angle=0]{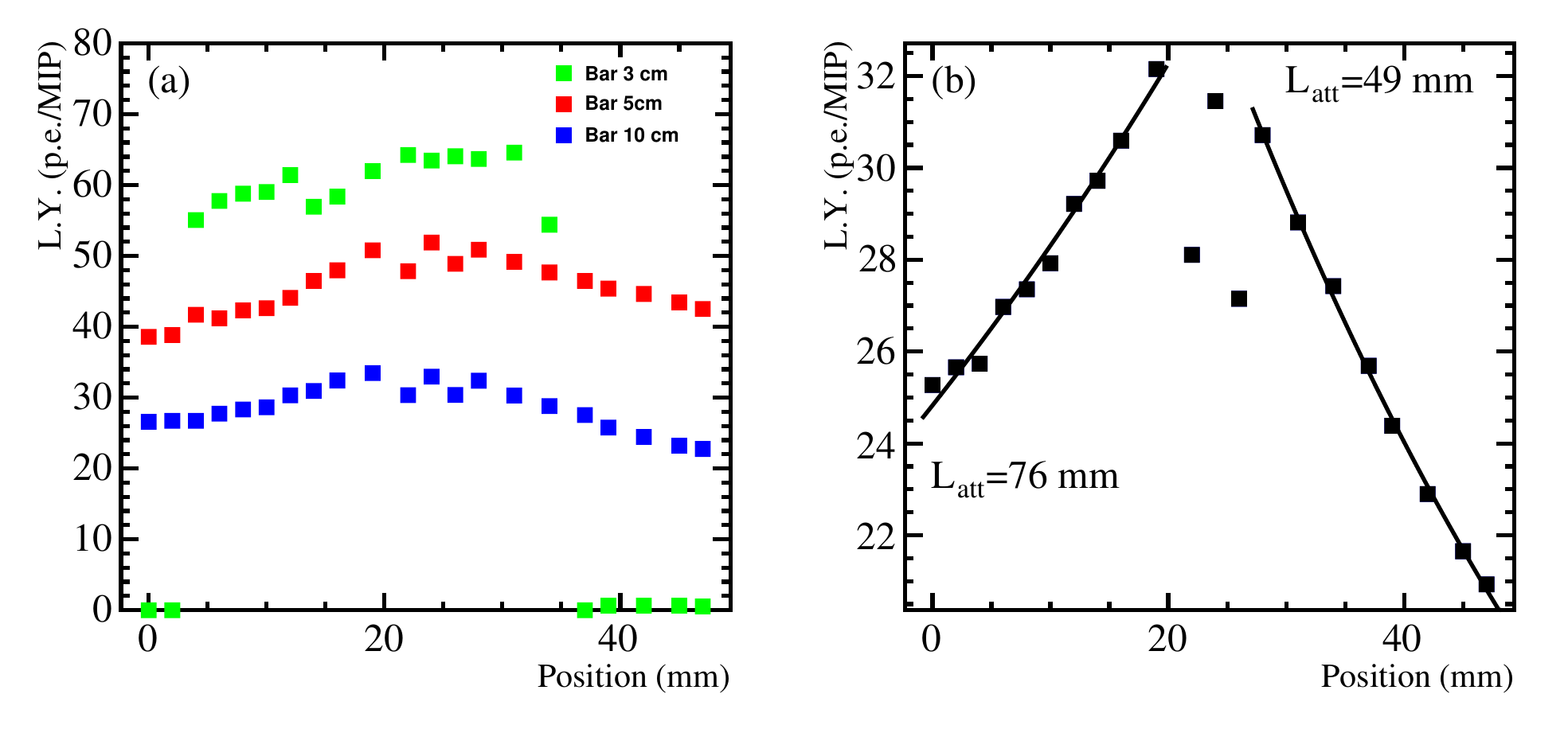}
\caption{\label{fig:ly_across}  Light yield scan across 3 cm, 5 cm and 10 cm bars (a). Attenuation of scintillating light in the opposite directions from a fibre  in the 10 cm bar (b).}
\end{center}
\end{figure}

The results are shown in Figure~\ref{fig:ly_across} (a).  The first point of the scan 
was arbitrary fixed near the bar edges. 
The scan spanned over 5~cm towards the opposite  edge crossing a fibre position at the coordinate of 
around 20--25~mm. The light yield is the sum of the light measured by the two photosensors at both ends of the bar. 
The attenuation of the scintillating light in the opposite directions from a fibre is demonstrated in Figure~\ref{fig:ly_across} 
(b) for the 10 cm wide bar.  Light attenuation is asymmetrical because of the effect of close reflective edge in one direction. 
The points were fitted with exponential function $f(x) = C\cdot exp(S\cdot x)$, where $x$ is the position variable and 
attenuation length is $1/S$. An attenuation length of 76~mm has been obtained towards the near edge, 
and 49~mm in the opposite direction, where the edges influence on the scintillating light collection is less important. 
The second value can be considered to a  first approximation  as the attenuation length of scintillating 
light propagation in 7~mm thick extruded scintillator.

\subsection{Detection efficiency}
\label{ssec:det-eff}

Data obtained during the scan measurements across the bars  have been used to calculate the detection efficiency.   
The electronic trigger signal was produced by two trigger counters in coincidence. 
An additional trigger counter of 3$\times$3~mm$^2$ active area allowed us to localize the position 
of the impinging beam on  the tested bars with high accuracy.
Only data collected with the beam impinging far from the bar edges have been considered for this measurement.
Events with the signals from the small counter with a pulse amplitude higher than 10~p.e. and time coordinate 
within $\pm \, 4 \sigma_t$ with respect to the average time
were selected for measuring the detection efficiency.

The detection efficiency has been obtained with three different methods: 
\begin{enumerate}
\item as the ratio between the number of hits within 4 $\sigma_t$ of the  $(t_{\rm L}-t_{\rm R})/2$ spectrum and the total number of triggers, 
where $t_{\rm L}$ and $t_{\rm R}$ are the times measured from the SiPMs situated at the left and right bar end with respect to the direction of the beam ({\it timing AND});
\item as  the ratio between the number of hits within 4$\sigma_t$ of the time distribution of one of the two photosensors and the total number of triggers ({\it timing OR});   
\item as the ratio between the number of hits with an integrated charge over some threshold at both bar ends and the total number of triggers ({\it charge AND}).  
\end{enumerate}

The {\it timing AND} method implies the most stringent selection criterion while the {\it timing OR} depends on the noise level of the photosensors.
The {\it charge AND} method  is the loosest criterion and it is  affected by  accidentals within 200~ns charge integration window. 

The results are listed in Table~\ref{table:eff} for the three methods and for 3 , 5 and 10 cm  wide bars.  The 10 cm wide bar 
was read out by a single fibre at both ends. Only signals with more than 3 p.e. at each bar end have been considered in the analysis.

\begin{table}[h]
\caption{Detection inefficiency  for different prototypes and for different methods of counting the detected events. 
 The uncertainty is purely statistical.}
 \begin{center}
 \begin{tabular}{|c|ccc|}  
 \hline
  Method of counting &  (1-$\varepsilon$) (\%)& (1-$\varepsilon$)  (\%) & (1-$\varepsilon$) (\%)\\
                               &   width = 3 cm         & width = 5 cm    & width = 10 cm \\ 
 \hline
 Timing AND    & 0.32$\pm$0.03   &    0.26$\pm$0.02   &    0.59$\pm$0.04    \\
 Timing OR      & 0.27$\pm$0.03    &   0.13$\pm$0.02   &    0.09$\pm$0.01   \\
 Charge AND   & 0.17$\pm$0.02   &    0.06$\pm$0.01  &    0.32$\pm$0.03    \\
 \hline
\end{tabular}
\end{center} 
\label{table:eff}  
\end{table}

\begin{figure}[htb]
\begin{center}
\includegraphics[width=10cm,angle=0]{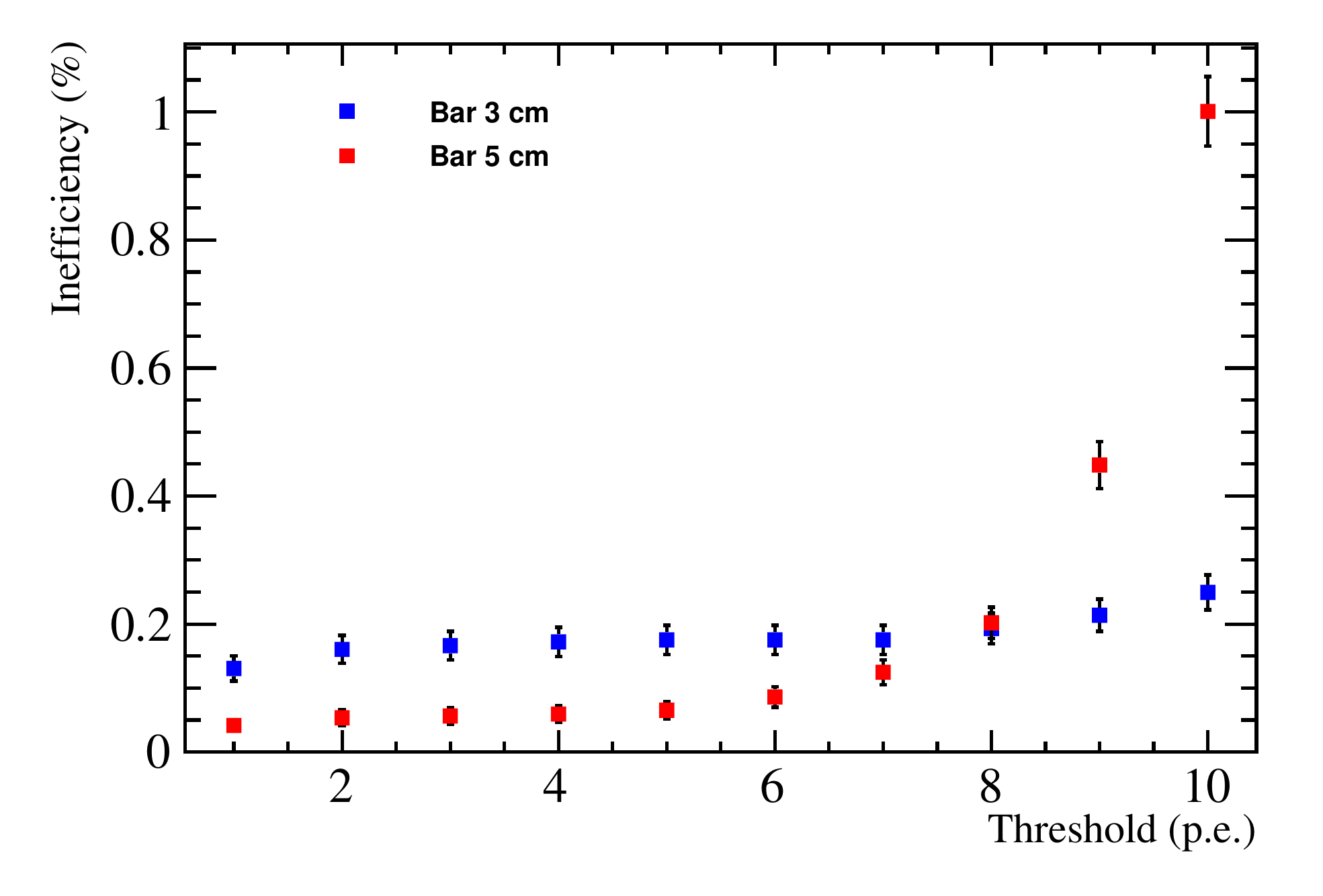}
\caption{\label{fig:ineff35}  The inefficiency for 3 and 5 cm wide bars as a function of the charge detection threshold. 
The events are counted if the charge is over the specified threshold at each bar end. }
\end{center}
\end{figure}

Figure~\ref{fig:ineff35} shows how the inefficiency depends on the  detection threshold in the case of {\it Charge AND}. The threshold is applied at each bar end.

\subsection{Time resolution}
\label{ssec:time-res}


The time resolution of bars L1, L2, and L4 was measured using the following procedure, repeated for each impact
position of the beam along the bar. 

\begin{figure}[htbp]
\centering 
\includegraphics[width=.6\textwidth]{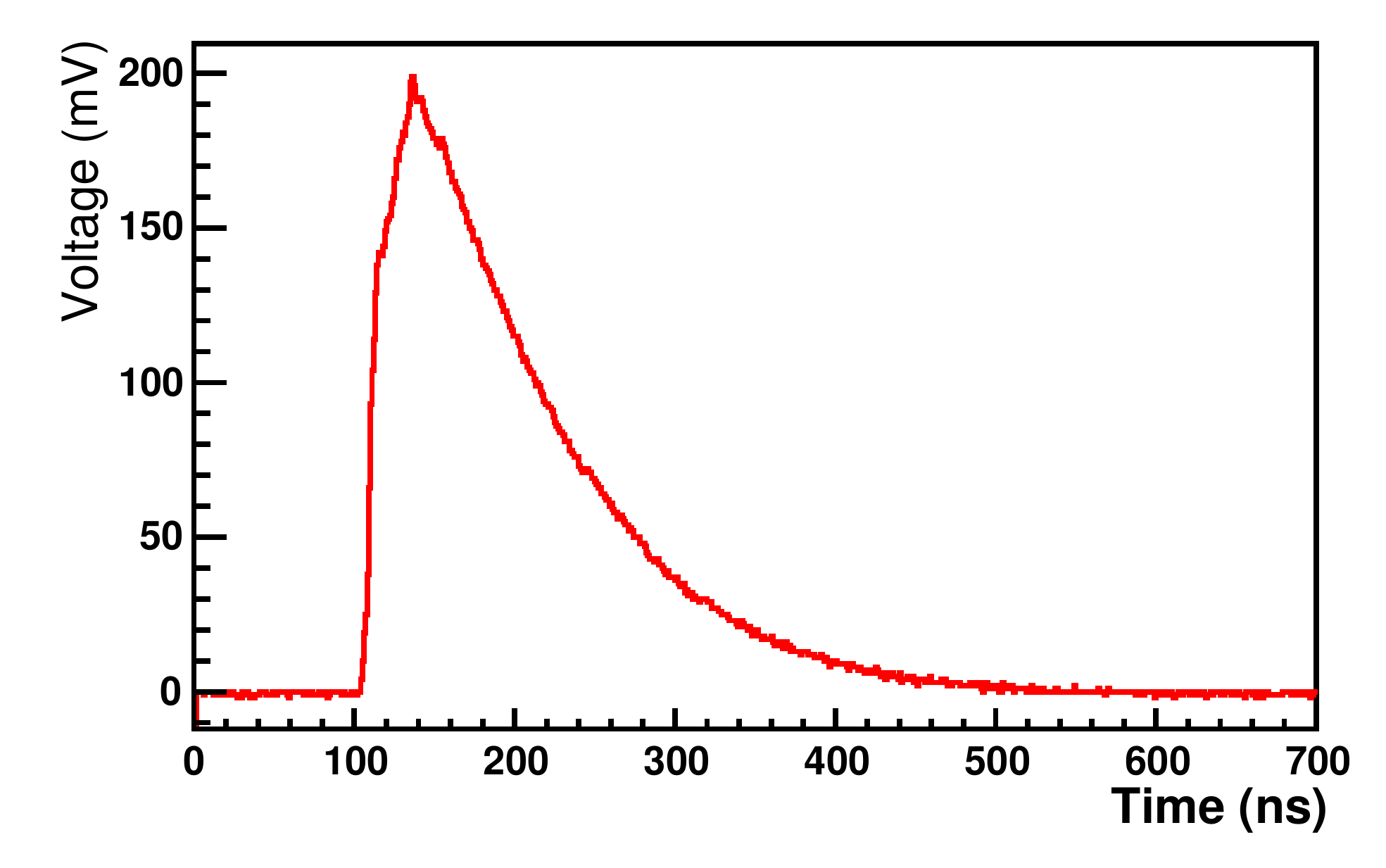}
\qquad
\caption{\label{fig:waveform} Example of baseline-subtracted waveform for 2 cm thick NICADD bars.}
\end{figure}

As a first step, the charge samples from the SiPMs ``left'' (L)  and ``right'' (R) with respect to the beam direction at the two bar ends were scanned in 0.5 ns steps, 
identifying a ``start'' time ($t_{\rm L}, t_{\rm R}$)
as the first time bin corresponding to an ADC count greated than 10 with respect to the average baseline.
An example of baseline-subtracted waveform is shown in Figure~\ref{fig:waveform}.
The same procedure has been applied to the trigger signals, obtaining a $t_0$ for each event 
that was subtracted from $t_{\rm L}$ and $t_{\rm R}$. 

As a second step, in order to compute the time-slewing corrections, the total charge $Q_{{\rm L,R}}$ distributions were calculated,
where $Q_{\rm L,R}$ are the charges integrated over 350 ns starting at $t'_{\rm L,R}=t_{\rm L,R}-t_0$.

The $Q_{\rm L,R}$ spectra were divided in 10 slices, each slice containing the same number of events, and for each slice a gaussian fit 
of the $t'_{\rm L,R}$ distribution was done.

\begin{figure}[htbp]
\centering 
\includegraphics[width=.6\textwidth]{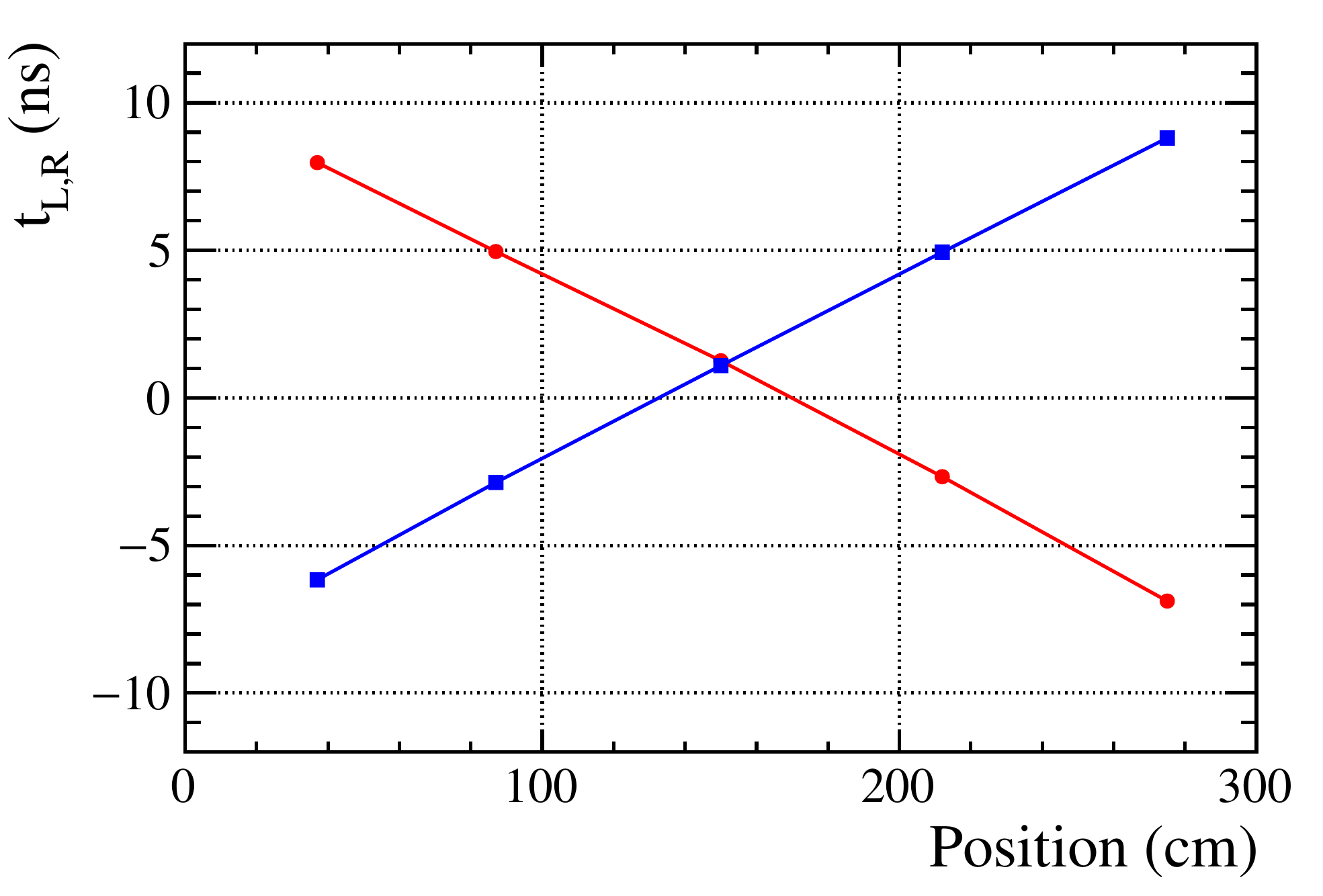}
\qquad
\caption{\label{fig:L2_v} Speed of light propagation $v_f$ in L2 fibre, for signals read by SiPM left (red line) and right (blue line); 
$v_f$ is measured as $(16.0\pm0.1)$ cm/ns.}
\end{figure}

The average values of these fits, together with the correponding values of $Q_{\rm L,R}$, were used to find by linear interpolation
a time-slewing correction term, as a function of $Q_{\rm L,R}$, to be subtracted from the previous $t'_{\rm L}$ and $t'_{\rm R}$,
obtaining new ``start'' values $t''_{\rm L,R} = t_{\rm L,R} - t_0 - t^{\rm TS}_{\rm L,R}$.
The values of $t''_{\rm L,R}$ were histogrammed and gaussian fits were remade, this time for all charges $Q_{\rm L,R}$ together. 
The averages of $t''_{\rm L,R}$ as a function of the beam impact position measure the speed of light propagation along the fibre, 
and is shown in Figure~\ref{fig:L2_v} for bar L2.

\begin{figure}[htbp]
\centering 
\includegraphics[width=.6\textwidth]{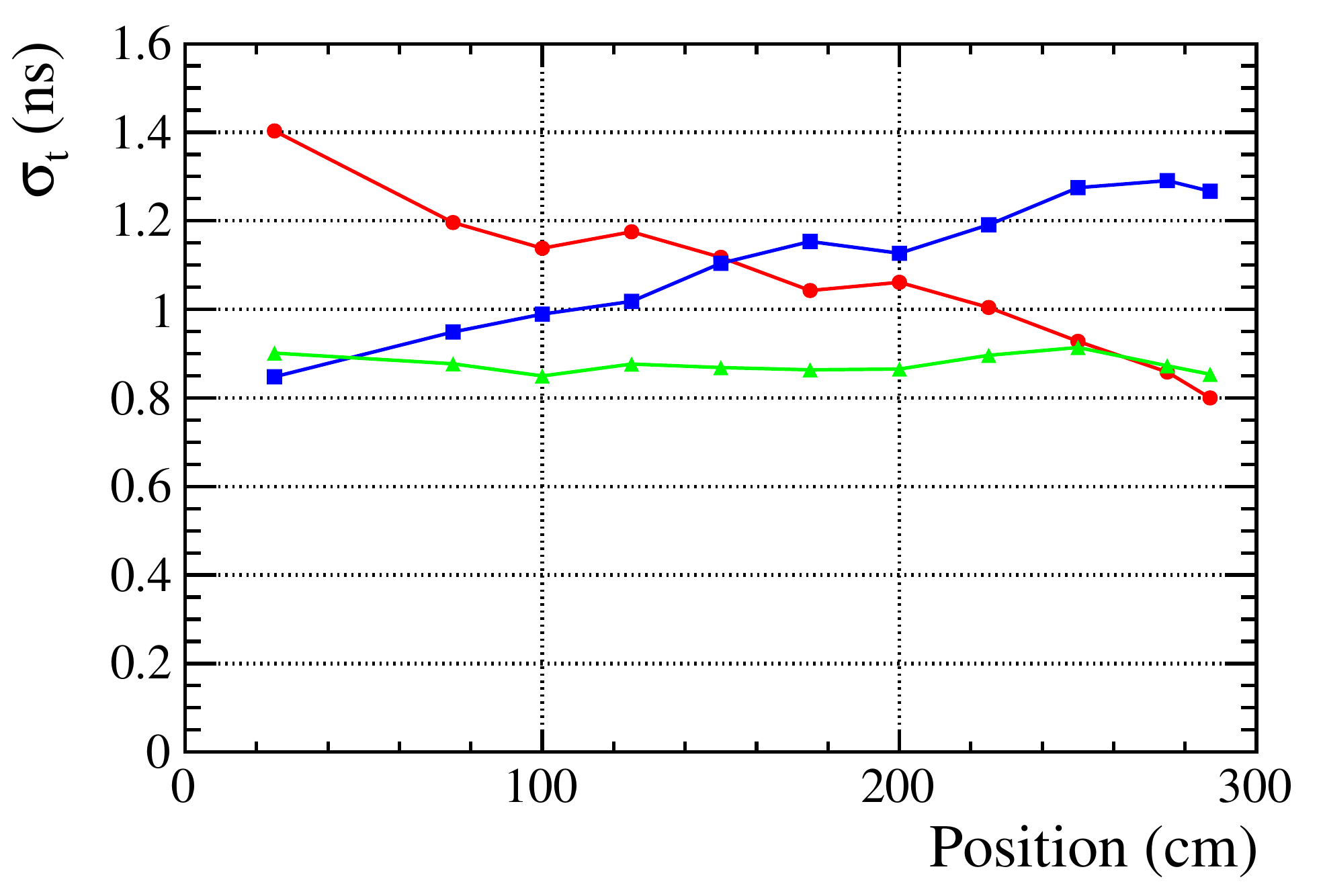}
\qquad
\caption{\label{fig:L1_s} L1 bar time resolution using only SiPM $L$($R$), red circles(blue squares) and both SiPMs (green triangles).}
\end{figure}

\begin{figure}[htbp]
\centering 
\includegraphics[width=.6\textwidth]{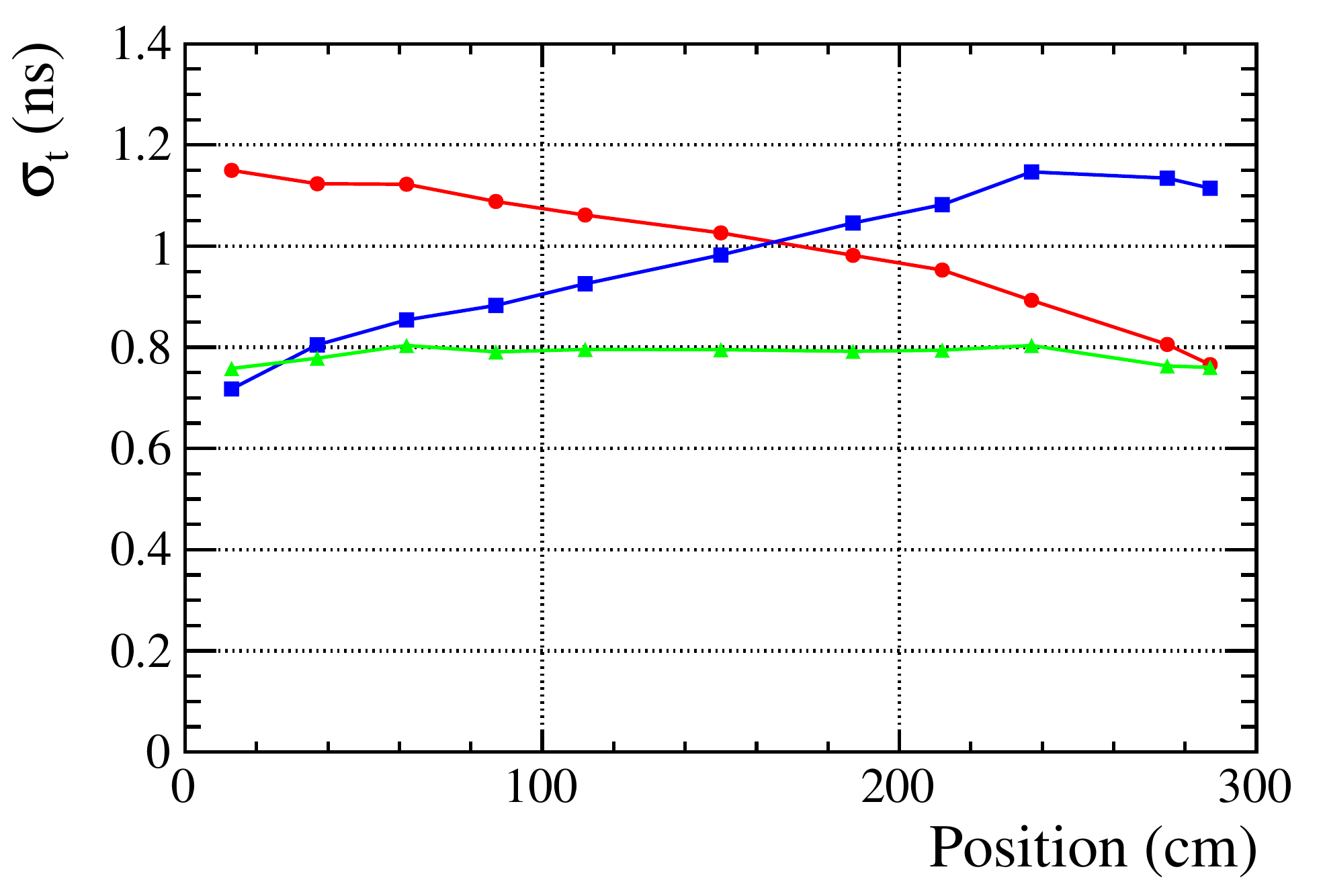}
\qquad
\caption{\label{fig:L2_s} L2 bar time resolution using only SiPM $L$($R$), red circles(blue squares) and both SiPMs (green triangles).}
\end{figure}

\begin{figure}[htbp]
\centering 
\includegraphics[width=.6\textwidth]{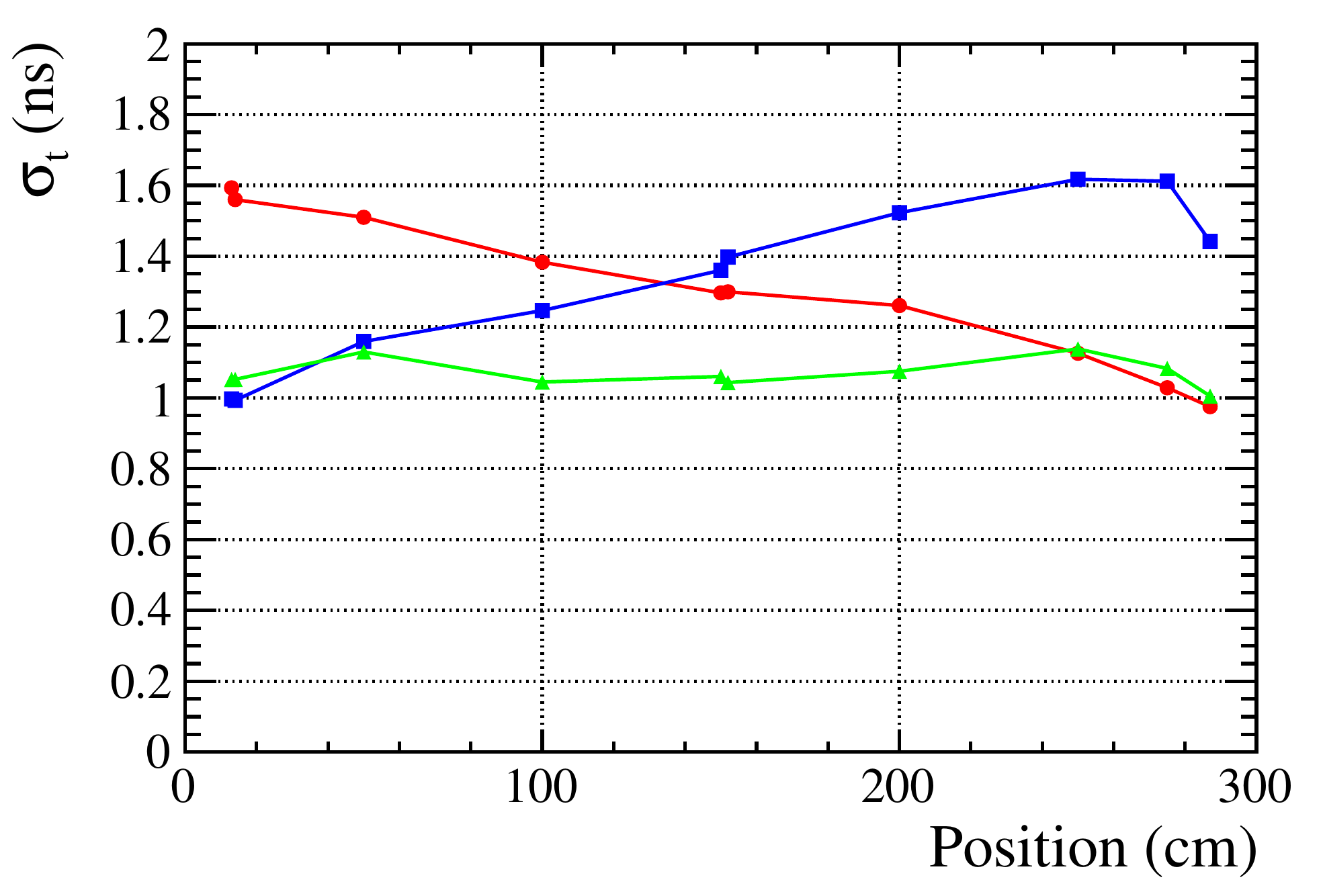}
\qquad
\caption{\label{fig:L4_s} L4 bar time resolution using only SiPM $L$($R$), red circles(blue squares) and both SiPMs (green triangles).}
\end{figure}

The time resolution for each bar as a function of the position of the beam along the fibre length was measured as the $\sigma$ of a Gaussian function
used to fit the time distributions of each SiPM individually and  the sum of the two, using the relation $0.5\cdot (t''_L + t''_R)$.
The results  are shown in Figures~\ref{fig:L1_s},~\ref{fig:L2_s},~\ref{fig:L4_s} respectively for bars L1, L2 and L4.

From the point of view of the time resolution $\sigma_t$, the L2 bar provides the best results. The results obtained from bar L1
would allow a great economy of scintillating material. For the L4 bar, with a 1.2 mm diameter fibre, the time resolution appears
very marginal. 

For completeness, we show in Table~\ref{tab:shortBars} the time resolution obtained in exactly the same way for short bars, read out only at one end,
and with the beam impact point at 13 cm far apart from the photosensor.

\begin{table}[htbp]
\centering
\caption{\label{tab:shortBars}Time resolution for short bars S1, S2, S4, S5 and S8 defined in Table~\ref{tab:prototypes1} when the beam 
impinges at $\sim$ 13 cm with respect to the SiPM. Treatment of slewing corrections is the same described in the text for long bars L1--L4.}
\smallskip
\begin{tabular}{|l|c|}
\hline
            &  time resolution [ns]   \\ \hline
S1          & $0.756 \pm 0.006$ \\
S2          & $0.676 \pm 0.005$ \\
S4          & $0.820 \pm 0.007$ \\
S5          & $0.676 \pm 0.005$ \\
S8          & $0.730 \pm 0.005$  \\ \hline
\hline
\end{tabular}
\end{table}

\begin{figure}[htb]
\begin{center}
\includegraphics[width=10cm,angle=0]{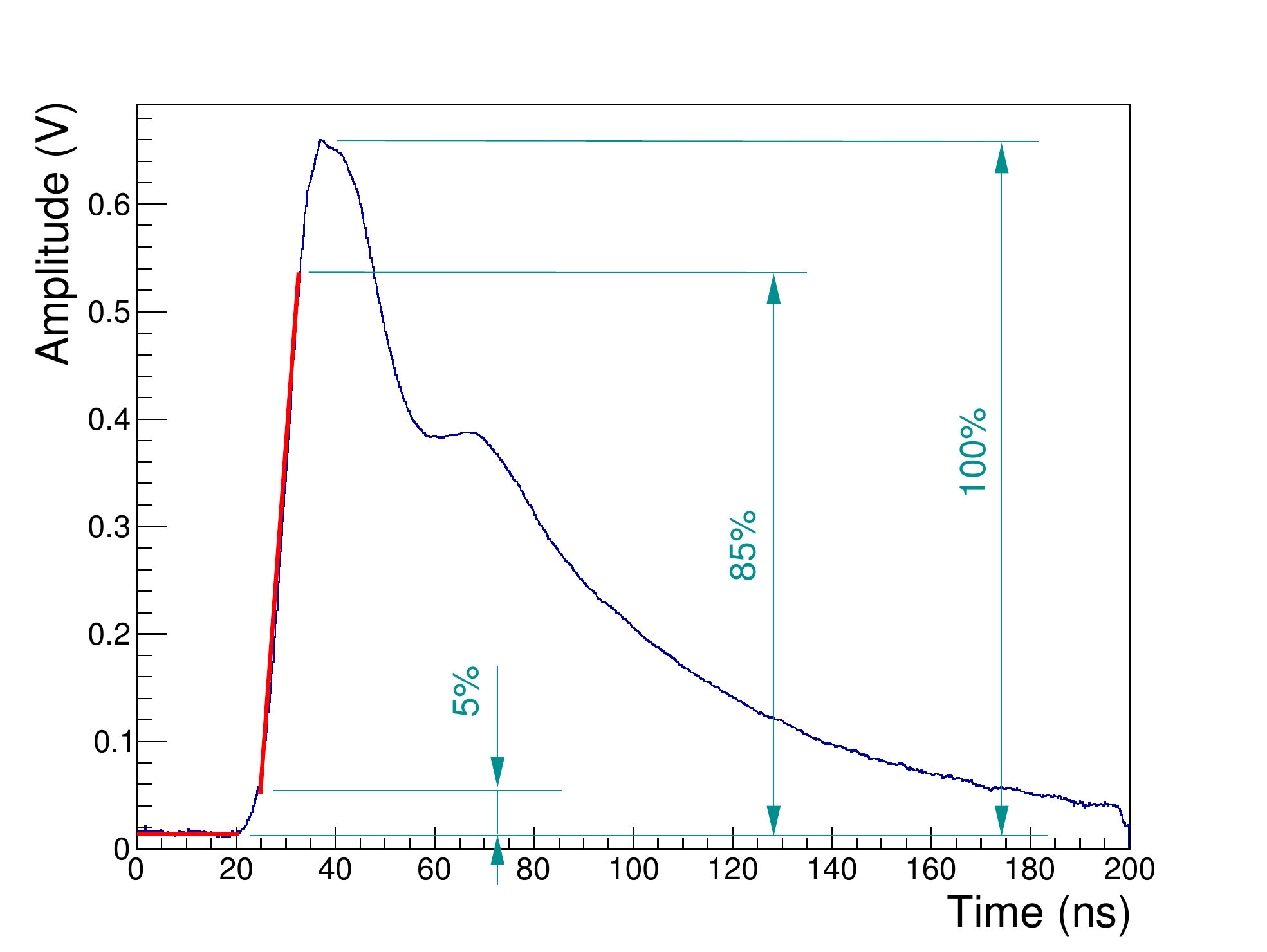}
\caption{ \label{fig:pulse_shape}  Fitting of digitized pulse shape to obtain the timing mark. }
\end{center}
\end{figure}

\vskip 2mm
The measurement of the time resolution for UNIPLAST bars has been performed as follows.
The waveform digitizer captures the pulse shape with steps of 200~ps each as shown in Figure~\ref{fig:pulse_shape}.
The typical pulse shape after the slow preamplifier is stretched over 1000 time samples, that correspond to 200~ns. 
The shoulder observed in the pulse is caused by the signal reflection in 2.5~m long twisted pair cable between MPPC and the preamplifier.
The rise time of the pulse was used  to obtain the timinig mark of the event.  The typical number of samples within the 
pulse front is over 60. This allows us to apply two different methods for calculating the timing, 
obtaining similar results.  


The first method consists in fitting with straight lines the rising edge of the pulse shape  
and the baseline before the signal;
the rising edge of the pulse shape is fitted between 5\% and 85\% of its maximum height.
The crossing point of the lines gives the relative time coordinate. The method is illustrated in  Figure~\ref{fig:pulse_shape}.

The time resolution was obtained by fitting with a Gaussian function the time distributions
$0.5 \cdot (t''_R + t''_L)$ at different positions of the beam along the bar length.

\begin{figure}[htb]
\begin{center}
\includegraphics[width=14cm,angle=0]{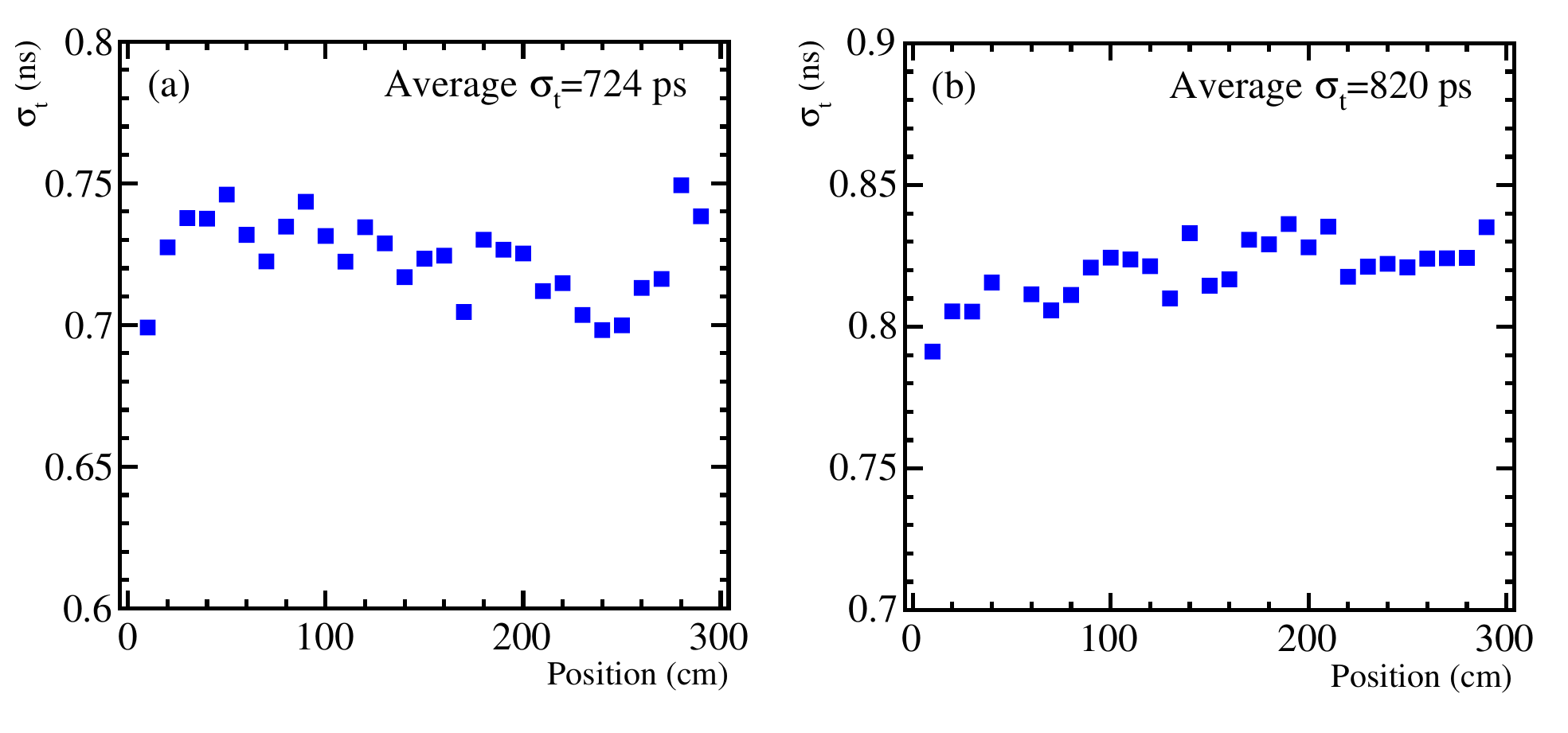}
\caption{\label{fig:time35}  Time resolution for 3 cm (a) and 5 cm bars (b) vs position along the bars. }
\end{center}
\end{figure}

The results  for 3 and 5 cm wide bars are shown in Figure~\ref{fig:time35}.  The points  are the average time resolution 
 for the three tested bars of the same size. The average resolutions along the whole bar length are 
$\sigma_t$ = 724~ps and $\sigma_t$=820~ps for 3 and 5 cm wide bars, respectively.
%

\begin{figure}[htb]
\begin{center}
\includegraphics[width=14cm,angle=0]{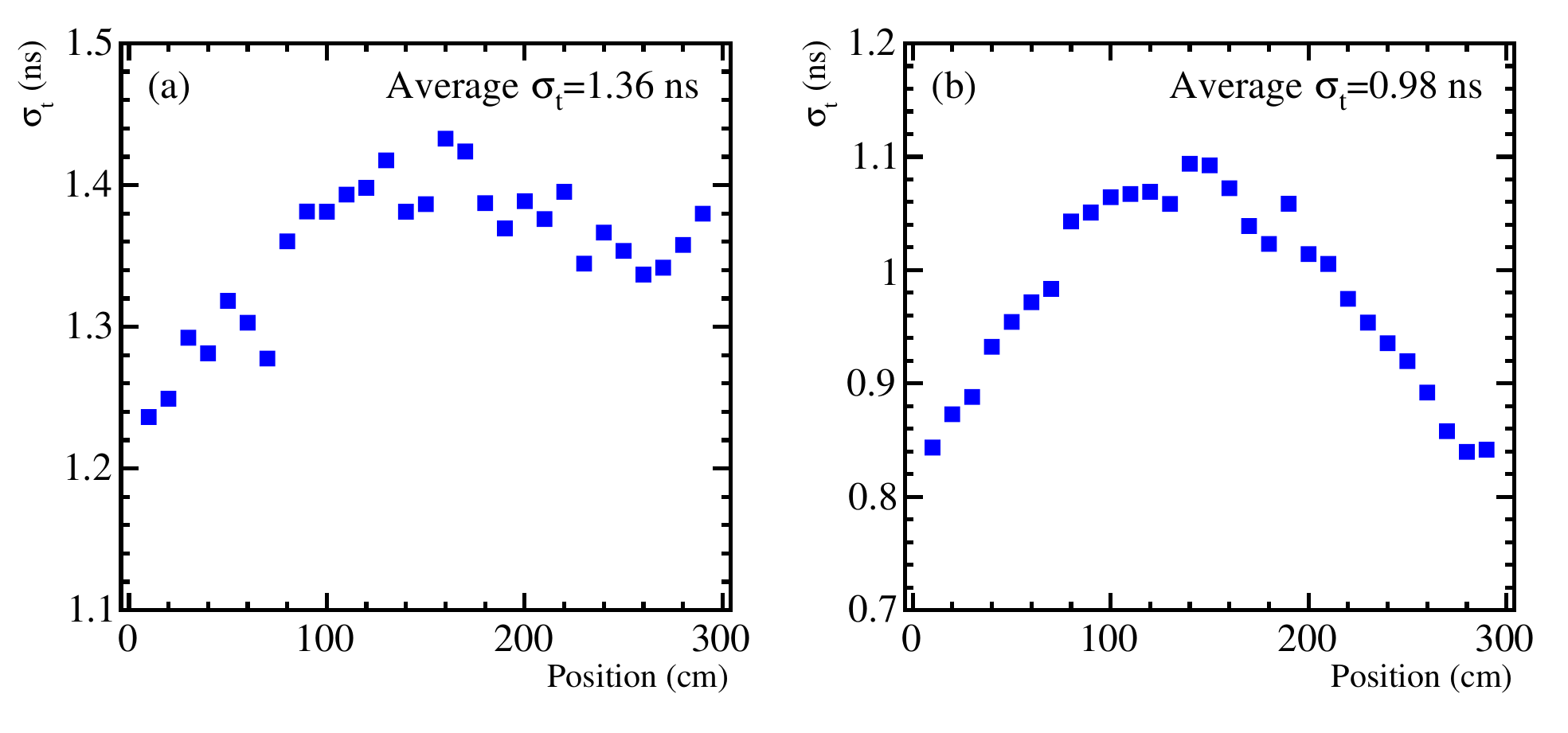}
\caption{\label{fig:time10}  Time resolution for 10 cm bars with 2 MPPC readout (a) and  4 MPPC readout (b). }
\end{center}
\end{figure}

The results for 10 cm bars are presented in Figure~\ref{fig:time10}~(a) for the readout  with two MPPCs (U4) 
and in Figure~\ref{fig:time10}~(b) for the  readout with four MPPCs (U3).  
The time resolution for the U4 and U3 bars were determined by fitting with a Gaussian function the distributions  $(t''_{\rm L} + t''_{\rm R}) \cdot 0.5$ and 
 $(t''_{\rm 1L}+t''_{\rm 1R} + t''_{2L} + t''_{2R}) \cdot 0.25$, respectively.
We find $\sigma_t = 1.4$ ns for the bar readout by two MPPCs (U4) and $\sigma_t = 1$ ns
for the bar read out by 4 MPPCs.
The time resolution of the bar instrumented with four photosensors improves by $\sqrt{2}$, as expected.

These results  were obtained for the whole  spectrum of pulse amplitudes. 
The time resolution of a single MPPC as a function of the light yield can be fitted as 
\[
\sigma_t=6.6 \; {\rm ns}/\sqrt{L.Y.(p.e.)}-0.14 \; {\rm ns}.
\]

The second method to determine the time resolution consists in simulating the behaviour of a constant fraction discriminator
on the pulse shape recorded by the digitizer.
The time mark was fixed to the clock sample where the pulse amplitude exceeds the  15\% fraction of the pulse height. 
The results were compatible with those obtained with the first method within the uncertainties.  

\section{Conclusions}
\label{sec:conclusions}
Parameters as the light yield,  time resolution and efficiency for minimum ionizing particles 
of different types of 300 cm and 25 cm long scintillating bars  from NICADD and UNIPLAST companies 
instrumented with wavelength 
shifting fibres and read out by  different models of silicon photomultipliers have been measured at a test 
beam at the T9 area at the CERN Proton Synchrotron.  

A time resolution of 700-800~ps constant along the bar length and a light yield of 140 (70) photoelectrons 
has been measured for 3~m long, 4.5 (5)~cm wide and
2 (0.7)~cm thick bars from NICADD (UNIPLAST) company.
The difference in light yield is due to the different scintillator properties, different bar geometry and different photosensors.
The detection efficiency for minimum ionizing particles exceeds 99.5\% for the prototypes from UNIPLAST company.

The results collected so far nicely match the requirements for  the SHiP muon detector.


\acknowledgments


The authors would like to thank H.~Wilkens, L.~Gatignon and M.~Jaeckel for continuous support and help.
This project has received funding from the European Union's  Horizon 2020 research and  innovation program under grant agreement No 654168.
This work has been supported by the Grant \#14-12-00560 of the Russian Science Foundation.


\end{document}